\documentclass[amsmath,amssymb,11pt]{article}
\usepackage{graphicx}
\usepackage{bm}
\usepackage{amsmath,amssymb}
\usepackage{color}
\usepackage{fullpage}%
\usepackage[affil-it]{authblk}
\usepackage{enumerate}
\usepackage[dvipdfmx,colorlinks,citecolor=blue]{hyperref}

\newtheorem{theo}{Theorem}
\newtheorem{defi}{Definition}
\newtheorem{lemm}{Lemma}
\newtheorem{remk}{Remark}

\begin{document}
\title{\bf
Commuting Quantum Circuits 
and
Complexity of 
Ising Partition Functions
}  
\author[1,2]{Keisuke Fujii}
\author[3]{Tomoyuki Morimae}

\affil[1]{The Hakubi Center for Advanced Research, Kyoto University}
\affil[ ]{Yoshida-Ushinomiya-cho, Sakyo-ku, Kyoto 606-8302, Japan}
\affil[2]{Graduate School of Informatics, Kyoto University}
\affil[ ]{Yoshida Honmachi, Sakyo-ku, Kyoto 606-8501, Japan}
\affil[3]{ASRLD Unit, Gunma University}
\affil[ ]{1-5-1 Tenjin-cho Kiryu-shi Gunma-ken, 376-0052 Japan}

\date{\today}
        \maketitle  
    
\begin{abstract}
Instantaneous quantum polynomial-time ({\sf IQP}) computation
is a class of quantum computation consisting only of 
commuting two-qubit gates and is not universal. 
Nevertheless, it has been shown that
if there is a classical algorithm that can simulate 
{\sf IQP} efficiently,  
the polynomial hierarchy ({\sf PH}) collapses to the third level,
which is highly implausible.
However, the origin of the classical intractability is 
still less understood.
Here we establish a relationship between 
{\sf IQP} and computational complexity of 
calculating the partition functions of Ising models.
We apply the established relationship
in two opposite directions.
One direction is
to find subclasses of {\sf IQP} that are classically efficiently simulatable
by using exact solvability of certain types of Ising models.
Another direction is 
applying quantum computational complexity of {\sf IQP}
to investigate (im)possibility of efficient classical
approximations of Ising partition functions with imaginary coupling constants.
Specifically, we show that
a multiplicative approximation of 
Ising partition functions 
is \#{\sf P}-hard for almost all imaginary coupling constants
even on planar lattices of a bounded degree.
\end{abstract}

\section{Introduction}
Quantum computation has a great possibility
to offer substantial advantages in solving 
some sorts of mathematical problems and also in simulating 
physical dynamics of quantum systems.
A representative instance is Shor's factoring algorithm~\cite{Shor},
which solves integer factoring problems in polynomial time,
while no polynomial-time classical algorithm has been known.
Recently, 
quantum algorithm for 
approximating Jones polynomial~\cite{AharonovJones,AharonovJonesHard}, Tutte polynomial~\cite{AharonovTutte},
and Ising partition functions~\cite{Cuevas11,Iblisdir,Matsuo14} have been found 
and they are shown to be {\sf BQP}-complete 
in certain parameter regions.
Furthermore, there are some evidences that
quantum computation, more precisely, {\sf BQP}
(bounded-error quantum polynomial-time computation~\cite{BernsteinVazirani}),
can solve problems outside the polynomial hierarchy ({\sf PH}\cite{PH1,PH2})
\cite{Aaronson09}.
These results strike the extended Church-Turing thesis~\cite{Turing,Church,BernsteinVazirani}, 
which states that every reasonable physical computing devices 
can be simulated efficiently (with a polynomial overhead) on a probabilistic Turing machine.
One of the most revolutionary and challenging goals
of human beings is to realize a universal quantum computer and
verify such quantum benefits in experiments.
However, experimental verification, 
which is the most essential part in science,
is still extremely hard to achieve,
requiring a huge number of qubits and extremely high accuracy in controls.

Is there any possible pathway to
verify computational complexity benefits of 
quantum systems that are realizable in the near future,
say, one-hundred-qubit (or particle) systems 
under reasonable accuracy of controls?
If there is such a subclass of quantum computation
that consists of experimental procedures much 
simpler than universal quantum computer 
but is still hard to simulate efficiently in classical computers,
experimental verification of 
complex quantum systems reaches a new phase. 

Aaronson and Arkhipov introduced {\sf B{\footnotesize OSON}S{\footnotesize AMPLING}}~\cite{boson},
a sampling problem according to the probability distribution of $n$ bosons
scattered by linear optical unitary operations.
The probability distribution
is given by the permanent of a complex matrix, 
which is determined by the linear optical unitary operations.
Calculation of the permanent of complex matrices
is known to be \#{\sf P}-hard~\cite{Valiant,AaronsonLinear}.
Since a polynomial-time machine with an oracle for \#{\sf P} can solve
all problems in the {\sf PH} according to Toda's theorem~\cite{Toda},
an exact classical simulation (in the strong sense~\cite{WeakSim,JozsaNest}
meaning a calculation of the probability distribution of the output)
of {\sf B{\footnotesize OSON}S{\footnotesize AMPLING}} is highly intractable 
in a classical computer.
They showed under assumptions of plausible conjectures that 
if there exists an efficient classical approximation of {\sf B{\footnotesize OSON}S{\footnotesize AMPLING}}
(classical simulation in the weak sense~\cite{WeakSim,JozsaNest}
meaning a sampling according the probability distribution of the output),
the {\sf PH} collapses to the third level, which unlikely occurs.
(The detailed notions of classical simulation are provided in Sec.~\ref{sec2}.)
This result brings a novel perspective on
linear optical quantum computation and 
drives many researchers into the recent proof-of-principle experiments
~\cite{BSex1,BSex2,BSex3,BSex4,BSex7,BSex8,BSex5,BSex6}.

Another subclass of quantum computation
of this kind
is instantaneous quantum polynomial-time computation ({\sf IQP})
proposed by Shepherd and Bremner~\cite{IQP0}.
{\sf IQP} consists
only of commuting unitary gates,
such as $\exp[ i \theta  \prod _{k \in S} Z_k]$.
Here $\theta \in [0,2\pi )$ is a rotational angle, $Z_k$ indicates the Pauli operator
on the $k$th qubit, and $S$ indicates a set of qubits
on which the commuting gate acts. (A detailed definition will be provided in the next section.)
The input is given by $|+\rangle ^{\otimes n}$ with $|+\rangle \equiv (|0\rangle + |1\rangle)/\sqrt{2}$,
and the output qubits are measured in the $X$-basis.
Since all unitary operations are commutable with each other, there is no temporal structure
in the circuits. (This is the reason why it is called 
instantaneous quantum polynomial-time computation.)
The commutability implies that
{\sf IQP} cannot perform an arbitrary unitary operation 
for the input qubits and hence
seems to be less powerful than
standard quantum computation, i.e., {\sf BQP}.
Nevertheless, Bremner, Jozsa, and Shepherd
showed that if there exists an efficient classical algorithm that
samples the outcomes according to
the probability distribution of {\sf IQP} with a certain multiplicative 
approximation error,
then the {\sf PH} collapses to the third level.
While the collapse of the {\sf PH} to the third level 
is not as unlikely as {\sf P} = {\sf NP},
it is also considered to be highly implausible.
This result is obtained by introducing postselection
and using the fact that post-{\sf BQP} = {\sf PP} shown by Aaronson~\cite{postBQP}.
Here postselection means that an additional ability to choose, without any computational cost, arbitrary measurement outcomes of possibly exponentially decreasing probabilities.
However, in comparison to {\sf B{\footnotesize OSON}S{\footnotesize AMPLING}}
~\cite{boson,Aaronson13},
the origin of the classical intractability of {\sf IQP}
is still not well understood.

\begin{figure}[htbp]
\begin{center}
\includegraphics[width=0.8\textwidth]{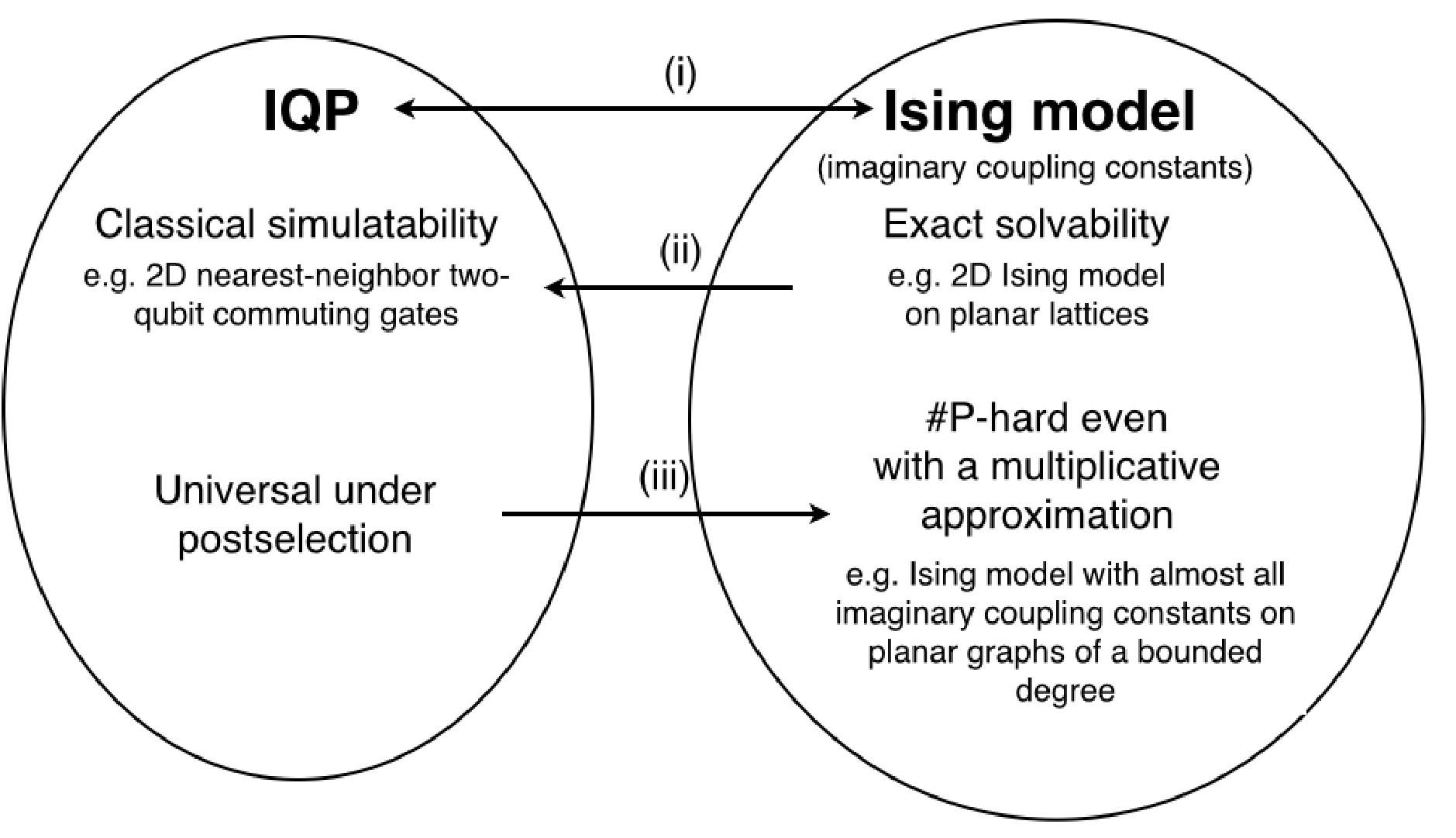}
\end{center}
\caption{The summary of results obtained in this work.} 
\label{fig0}
\end{figure}
The purpose of this paper is to further explore {\sf IQP}
by relating it with computational complexity of calculating Ising partition 
functions, which has been well studied in
statistical physics, condensed matter physics, and computer science.

Specifically we obtain the following results (see Fig.~\ref{fig0}):
\begin{enumerate}[(i)]
\item
We reformulate {\sf IQP} from a viewpoint of computational 
complexity of calculating Ising partition functions. 
The probability distribution of the output of {\sf IQP}
including its marginal distributions is  
mapped into an Ising partition function 
with imaginary coupling constants (Theorem~\ref{Main1} and Theorem~\ref{Main2}).

\item 
By using the above relation,
we specify classically simulatable classes of 
{\sf IQP}, which correspond to exactly solvable 
Ising models (Theorem~\ref{sparseIQP} and Theorem~\ref{freefermion}).
For example, 
{\sf IQP} that consists only of nearest-neighbor two-qubit commuting gates 
in two dimensions (2D) is 
classically simulatable, at least in the weak sense, irrespective of 
their rotational angles.

\item 
We show that
a multiplicative approximation of 
the Ising partition functions with almost all imaginary coupling constants
is \#{\sf P}-hard 
even on 2D planar lattices with a bounded degree.
So there is no polynomial-time approximation scheme 
unless the {\sf PH} collapses completely.
\end{enumerate}

The first result bridges 
{\sf IQP} and computational complexity of Ising partition 
functions,
which tells us the origin of 
hardness of classical simulation of {\sf IQP},
since exact calculation of Ising partition functions
is \#{\sf P}-hard even in the ferromagnetic case
\cite{Barahona,Jerrum93}.
Only restricted models are known to be exactly 
solvable such as Ising models on the 2D planar lattices
without magnetic fields.

One might naively expect that 
a subclass of {\sf IQP}, which is mapped into an exactly solvable 
Ising model, is classically simulatable in the strong sense~\cite{WeakSim,JozsaNest},
since the joint probability distribution of the output can be calculated efficiently.
However, there are exponentially many instances of 
the measurement outcome,
and hence an efficient calculation of
the joint probability distribution of an output does not directly applied to 
an efficient weak simulation of {\sf IQP}.
For example, in Ref.~\cite{WeakSim},
it is pointed out that
there exists the case where the joint probability distribution 
is easily calculated but its marginals are rather hard to calculate.
In order to construct an efficient weak simulation of {\sf IQP},
we need the marginal distributions,
which allow a recursive simulation of the sampling problem
by using the Bayes theorem.
To this end, we map not only the joint probability distribution
but also the marginal distributions
of {\sf IQP} into the Ising partition functions on another lattices.
In the proof, we virtually utilize measurement-based quantum 
computation (MBQC)~\cite{MBQC} on graph states~\cite{GraphState},
which are defined associated with the {\sf IQP} circuits.

The established relationship 
between {\sf IQP} and Ising partition functions 
is useful since computational complexity of Ising models have been well studied.
We can apply preexisting knowledge 
to understand quantum computational complexity of {\sf IQP}.
Specifically, in the second result,
we provide classical simulatable classes of 
{\sf IQP} by using exact solvability of certain types of Ising models.
We provide two examples of classically simulatable classes of {\sf IQP}.
One is based on the sparsity of the commuting gates.
Another is a class of {\sf IQP}
that consists only of two-qubit commuting gates
acting on nearest-neighbor qubits on the 2D planar graphs,
which we call planar-{\sf IQP}.
Planar-{\sf IQP} is mapped into
a two-body Ising model on a 2D planar lattice without magnetic fields,
which is known to be solvable by using the Pfaffian method~\cite{Barahona,Kasteleyn,Fisher}.
In the proof, we also utilize properties of graph states
in order to renormalize random $i\pi/2$ magnetic fields into two-body interactions,
which originated from the random nature of the measurements.
Then the marginal distributions can be 
efficiently calculated irrespective of 
their rotational angles by using the Pfaffian method 
\cite{Barahona,Kasteleyn,Fisher}.

On the other hand, {\sf IQP} consisting of 
single- and two-qubit commuting gates acting on a 2D planar graph
is sufficient to simulate universal quantum computation under postselection
~\cite{IQP}.
(Hereafter, such a property that a quantum computational task $A$
can simulate universal quantum computation under postselection is called 
as {\it universal-under-postselection}.)
This fact and the above classically simulatable class imply that single-qubit rotations 
play a very important role for {\sf IQP} to be classically intractable.
Actually single-qubit rotations make a drastic change 
of complexity from almost strongly simulatable to
not simulatable even in the weak sense.
A similar result is also obtained for 
Toffoli-Diagonal circuits, where 
the Hadamard gates at the final round
plays very important role~\cite{WeakSim}.

When experimentalists 
utilize {\sf IQP} for the purpose of 
verification of quantum benefits in malicious experimental setups,
they should avoid the {\sf IQP} circuits of these classically simulatable classes,
since a malicious experimental device
can cheat experimentalists by a classical sampling instead of implementing the {\sf IQP} circuits.
In most cases, however, experimental setups 
are well organized and not so malicious.
Thus it might be possible to use these classically simulatable classes 
combined with any physically relevant assumption
as an efficient benchmark of commuting quantum circuits,
since the ideal output distribution can be efficiently calculated.
Since planar commuting circuits 
can generate an interesting class of entangled states 
called weighted graph states~\cite{GraphState},
the constructed efficient classical simulation 
would be useful for an experimental verification of 
the weighted graph states.

In the above classically simulatable class,
the probability distribution is 
given by the determinant, i.e., square of the Pfaffian, 
of a complex matrix.
This result contrasts with 
{\sf B{\footnotesize OSON}S{\footnotesize AMPLING}} related with
the permanent of a complex matrix.
The exact solvability with the determinant (Pfaffian) naturally reminds us free-fermionic models,
which have been also studied in standard quantum computation
as match gates~\cite{matchgates0,matchgates1,matchgates2,matchgates3,matchgates4}.
Since a determinant can be mapped into 
a probability amplitude of a free-fermionic system,
the classically simulatable class of {\sf IQP}
can be regarded as {\sf F{\footnotesize ERMION}S{\footnotesize AMPLING}} discussed in Ref.~\cite{Aaronson13}.
This suggests that the sampling problems
in physics can be classified in a unified way
as sampling problems of elementary particles.

In the final result,
we apply the first result
in an opposite direction, from quantum complexity to classical one.
We consider certain universal-under-postselection instances of {\sf IQP}
to understand classical complexity of calculating 
the Ising partition functions.
Specifically we show that
a multiplicative approximation of
Ising partition functions (corresponding to a strong simulation of {\sf IQP} with 
a multiplicative error) 
is \#{\sf P}-hard
for almost all imaginary coupling constants 
even on 2D planar lattices with a bounded degree.
Hence if there exists a fully polynomial-time classical approximation scheme,
it results in a complete collapse of the {\sf PH}.
This can be viewed as a ``quantum proof" of \#{\sf P}-hardness of 
approximating the imaginary Ising partition functions.
Aaronson's post-{\sf BQP} = {\sf PP} theorem~\cite{postBQP},
which is employed to show the above result,
is also utilized to provide a ``quantum proof"~\cite{QuantumProof} of \#{\sf P}-hardness of 
approximating the permanent~\cite{AaronsonLinear} and the Jones polynomial~\cite{Kuperberg}
with a multiplicative error.

The rest of the paper is organized as follows.
In Sec.~\ref{sec2}, we introduce 
the definition and useful properties of the graph states
in order to fix the notation.
Then we review {\sf IQP} and the postselection argument
introduced by Bremner, Jozsa, and Shepherd.
We also mention how to utilize post-{\sf BQP} = {\sf PP} theorem by Aaronson~\cite{postBQP}
to obtain classical complexity results.
As the final part of the preliminary section, we summarize related works on commuting quantum circuits
and quantum and classical computational complexity of calculating the Ising partition functions.
In Sec.~\ref{sec3}, we establish a relationship 
between {\sf IQP} and Ising partition functions,
not only for the joint probability distribution of the output
but also for its marginal distributions.
In Sec.~\ref{sec4}, we demonstrate two classically simulatable classes 
of {\sf IQP}. 
One is based on the sparsity of 
the {\sf IQP} circuits. Another is based on
exact solvability of the Ising models on the 2D planar lattice 
without magnetic fields.
In Sec.~\ref{sec5}, we apply the relationship between
{\sf IQP} and Ising partition functions
in an opposite direction
to investigate (im)possibility of an efficient classical
approximation scheme of the Ising partition functions
with imaginary coupling constants.
Section ~\ref{sec6} is devoted to conclusion and discussion.

\section{Preliminary}
\label{sec2}
In the proofs of the main theorems,
we work with a measurement-based version 
of {\sf IQP}, namely {\sf MBIQP}, introduced by Hoban {\it et al.}~\cite{Hoban}.
The reason is that transformations on
the resource state for MBQC~\cite{MBQC}, 
so-called graph states~\cite{GraphState}, are much easier 
and more intuitive than transformations on the unitary gates themselves.
Here we introduce the definition and useful properties of graph states
in order to fix the notations.

\subsection{Basic Notations}
The Pauli matrix on the $i$th qubit
is denoted by $A_i$ ($A=I,X,Y,Z$).
The Hadamard gate is denoted $H$.
The eigenstates of $Z$ with eigenvalues $+1$ and $-1$ are denoted by $|0\rangle$ and $|1\rangle$, respectively.
The eigenstates of $X$ with eigenvalues $+1$ and $-1$ are denoted by $|+\rangle$ and $|-\rangle$, respectively.
We denote the controlled-$A$ gate acting 
on the $i$th (control) and $j$th (target) qubits by
$\Lambda _{i,j}(A) = |0\rangle \langle 0| \otimes I + |1\rangle \langle 1|  \otimes A$.
Specifically, $\Lambda _{i,j}(Z)=\Lambda _{j,i}(Z)$ and 
$H_j\Lambda _{i,j} (Z) H_j = \Lambda _{i,j}(X)$.

\subsection{Graph states}
\begin{defi}[Graph state]
\label{graphstate}
Suppose $G=(V,E)$ is a graph 
consisting of vertices $V$ and edges $E$.
We define the neighbor $\mathcal{N}_i$ of $i$ as the set of vertices 
adjacent to vertex $i$.
An operator $K_i= X_i \prod _{j \in \mathcal{N}_i}Z_j $ 
is defined for each vertex $i$.
The graph state $|G\rangle$ is defined as 
the simultaneous eigenstate of 
the operator $K_i$ with eigenvalue $+1$ for all $i$:
\begin{eqnarray*}
K_i |G\rangle = |G\rangle.
\end{eqnarray*}
\end{defi}
The above relation reads that
the graph state $|G\rangle$ is stabilized by
the operator $K_i$ for all $i$.
Such a state is called a stabilizer state.
The operator $K_i$, which stabilizes the stabilizer state,
is called a stabilizer operator.
A detailed description of the stabilizer formalism could be found in Refs.~\cite{Gottesman,NielsenChuang}.

The graph state $|G\rangle$ is generated 
from a tensor product state of 
$|+\rangle$
by performing $\Lambda _{i,j}(Z)$
on the pairs of qubits connected by edges $(i,j) \in E$:
\begin{eqnarray*}
|G\rangle = \left( \prod _{(i,j)\in E} \Lambda_{i,j} (Z) \right) |+\rangle^{ \otimes |V|}.
\end{eqnarray*}
This can be confirmed as follows. The product state $|+\rangle ^{ \otimes |V|}$
is the eigenstate of $X_i$ with eigenvalue $+1$ for all $i \in V$,
and hence $X_i | + \rangle ^{ \otimes |V|}= |+ \rangle ^{ \otimes |V|}$.
By applying $\prod _{(i,j)\in E} \Lambda_{i,j} (Z)$
for both sides, we obtain
\begin{eqnarray*}
\left( \prod _{(i,j)\in E} \Lambda_{i,j} (Z) \right) X_i | + \rangle ^{ \otimes |V|} &=& \left( \prod _{(i,j)\in E} \Lambda_{i,j} (Z) \right)|+ \rangle ^{ \otimes |V|}
\\
\Leftrightarrow
 K_i \left( \prod _{(i,j)\in E} \Lambda_{i,j} (Z) \right)| + \rangle ^{ \otimes |V|} &=& \left( \prod _{(i,j)\in E} \Lambda_{i,j} (Z) \right)|+ \rangle ^{ \otimes |V|},
\end{eqnarray*}
where we used the fact that $\Lambda _{i,j}(Z) X_i = X_i Z_j \Lambda _{i,j}(Z)$.
This is the definition of the graph state, and we conclude $|G\rangle = \left( \prod _{(i,j)\in E} \Lambda_{i,j} (Z) \right)|+ \rangle ^{ \otimes |V|}$.

In the proofs of the main theorems,
we repeatedly consider projective measurements on
the graph state and the resultant post-measurement graph state.
In the following we will see two important 
transformations on the graph states by 
projective measurements in certain bases.

\begin{remk}[$Z$-basis measurement]
\label{Zmeasurement}
If the $k$th qubit of the graph state $|G\rangle$ 
is measured in the $Z$-basis,
the resultant post-measurement state is 
the graph state associated with the graph $G' \equiv G \backslash k$,
where the byproduct operator $B_k=\prod _{j \in \mathcal{N}_k} Z_j$  
is located according to the measurement outcome $m_k \in \{ 0,1\}$,
i.e., $B_k ^{m_k} |G'\rangle$.
\end{remk}
{\it Proof}:
We observe the effect of the measurement on the stabilizer 
operator $K_i$.
If $i\neq k$ nor $i \neq \mathcal{N}_k$,
the measurement does not make any effect on a stabilizer $K_k$,
and hence the post-measurement state is stabilized by such a $K_k$.
If $i=k$, $K_i$ anticommutes with $Z_k$
and hence does not stabilize the post-measurement state anymore.
Instead, $(-1)^{m_k}Z_k$ stabilizes the post-measurement state
$|m_k\rangle _k$, where $m_k=0,1$ is the measurement outcome.
If $i\in \mathcal{N}_k$,
we define a new stabilizer operator $K'_{i} = Z_k K_i$
such that $K_k$ does not contain $Z_k$.
The post-measurement state is stabilized by $(-1)^{m_k}K'_{i}$.
Thus the graph state with the byproduct operator, $B_k^{m_k}|G' \rangle$,
is the post-measurement state.
(Note that $B_k^{m_k}$ anticommutes with $K'_i$s for all $i$
but commutes with $K_i$s with $i \neq k$ and $i \notin \mathcal{N}$.)
\hfill $\square$

Intuitively, the $Z$-basis measurement on the $k$th qubit removes
the $k$th qubit from the graph state,
and then the byproduct operator $B_k$ is located 
according the measurement outcome $m_k$.

Next we consider a projective measurement 
on the $k$th qubit
in the $\{|\theta _{k,m_k} \rangle \equiv X^{m_k} (e^{ - i \theta _k} |+\rangle + e^{ i \theta _k } |- \rangle )/\sqrt{2} \}$ basis,
where $m_k \in \{0,1\}$ is the measurement outcome.

\begin{remk}[Remote $Z$-rotation]
\label{remote-Z}
The projective measurement of the $k$th qubit
on the graph state $|G\rangle$ in the $\{ | \theta _{k,m_k} \rangle \}$
basis results in 
\begin{eqnarray*}
\exp\left[  i (\theta _k+ m_k\pi /2) \left( \prod _{j \in \mathcal{N}_k} Z_j \right)
\right] | G\backslash k \rangle/\sqrt{2}.
\end{eqnarray*} 
\end{remk}
{\it Proof}:
By using the fact that 
\begin{eqnarray*}
|G\rangle = \left(\prod _{j \in \mathcal{N}_k} \Lambda _{k,j} (Z) \right) |+\rangle _k |G \backslash k\rangle ,
\end{eqnarray*}
we can calculate the projection as follows:
\begin{eqnarray*}
\langle \theta _{k,m_k } |_k| G\rangle 
&=& \langle \theta _{k,m_k } |_k\left(\prod _{j \in \mathcal{N}_k} \Lambda (Z)_{kj} \right) |+\rangle _k| G \backslash k \rangle  
\\
&=&\langle +|_k e^{ i( \theta _k  + m_k \pi/2) Z_k } H_k  \left(\prod _{j \in \mathcal{N}_k} \Lambda (Z)_{kj} \right) |+\rangle _k| G \backslash k \rangle  
\\
&=& \left[\cos(\theta _k  + m_k \pi/2) I +  i \sin (\theta _k  + m_k\pi/2)\left(\prod _{j \in \mathcal{N}_k} Z_j \right) \right] | G \backslash k \rangle /\sqrt{2}  
\\
&=&
\exp\left[  i (\theta _k+m_k \pi /2) \left( \prod _{j \in \mathcal{N}_k} Z_j \right)
\right] | G\backslash k \rangle/\sqrt{2}.
\end{eqnarray*}
\\
\hfill $\square$.

The measurement in the $\{ | \theta _{k,m_k} \rangle \}$ 
basis induces a multi-body $Z$ rotation 
on the qubits adjacent to the $k$th qubit.
The norms of the post-measurement states are both 
$1/2$, which indicates that the outcomes $m_k=0,1$ 
appear randomly.

Another class of measurements, 
which is frequently used in MBQC, 
is the measurement 
in a $\{ e^{i \theta Z} |\pm \rangle \}$ basis.
It is known that 
adaptive measurements in these bases on a certain graph state
is enough to perform universal quantum computation, i.e., {\sf BQP}~\cite{MBQC}. 
Here the adaptive measurement means 
to change the following measurement angles 
according to the previous measurement outcomes
in order to handle the random nature of the measurements.
This process is often called a feedforward.
A wide variety of graph states 
have been known to be universal resources for MBQC~\cite{GraphState}.

\subsection{Strong and weak simulations of quantum circuits}
Here we provide definitions of two important notions for classical simulation of 
quantum circuits, strong and weak simulations~\cite{WeakSim,JozsaNest}.

\begin{defi}[Strong and weak simulations]
Suppose $\mathcal{C}$ is a uniformly generated quantum circuit of a 
model of quantum computation $A$ (e.g., {\sf IQP}, one-clean-qubit model~\cite{DQC1},  and universal quantum computation, etc.).
The probability distribution of the output $x$ (classical bits) 
is denoted by $P_{A}(x|\mathcal{C})$.
An efficient weak simulation of $A$ is a classical polynomial-time randomized computation
that samples $x$ with the probability $P_{A}(x|\mathcal{C})$.

On the other hand,
an efficient strong simulation of a quantum circuit $\mathcal{C}$ 
for a given output $x$ is 
a classical polynomial-time (randomized) computation that
calculates the probability $P_{A}(x|\mathcal{C})$
including its marginal distributions $\sum _{x'} P_{A}(x|\mathcal{C})$
with respect to an arbitrary subset $x'$ of the output bits $x$.
\end{defi}

In addition to these notions of classical simulation,
we can further consider types of approximations.
In an approximated simulation with a multiplicative error $1<c$,
we can replace the probability distribution $P_{A}(x|\mathcal{C})$ with
its approximation $P_{A}^{\rm ap}(x|\mathcal{C})$ that lies inside 
the following approximation range
\begin{eqnarray*}
\frac{1}{c} P_{A}(x|\mathcal{C}) \leq P^{\rm ap} _{A}(x|\mathcal{C}) \leq c P_{A}(x|\mathcal{C}).
\end{eqnarray*}

Apparently, if we can simulate $A$ in the strong sense,
we can sample the output in the weak sense.
Thus a strong simulation trivially includes a 
weak one. In fact, it has been known that
a strong simulation is much harder than a weak simulation, i.e., what a model of quantum computation $A$ can actually do.
For example,
an exact strong simulation of the output of 
universal quantum computation is \#{\sf P}-hard~\cite{WeakSim}.
We should also note that, 
in strong simulation, 
calculation of the marginal distributions is crucial,
since there is the case where
a strong simulation of the output probability (joint probability)
is easy but its marginal distributions are hard to calculate~\cite{WeakSim}.

\subsection{Instantaneous quantum polynomial-time computation}
Here we introduce {\sf IQP} and its measurement-based version.
We first define {\sf IQP}:
\begin{defi}[{\sf IQP} by Bremner et al.~\cite{IQP0,IQP}]
Let $n$ be the number of qubits.
A commuting gate is defined by
\begin{eqnarray*}
D(\theta _j,S_j)\equiv
\exp\Big[i\theta_j\prod_{k \in S_j }Z_k\Big],
\end{eqnarray*} 
where $\theta_j \in [0, 2 \pi)$ is a real number meaning the rotational angle,
and $\{ S_j\}$ is a set of subsets of $\{ 1,2,...n\}$, on which
the commuting gates act.
We refer to a poly($n$) number of commuting gates,
including the input state $|+\rangle ^{\otimes n}$ 
and the $X$-basis measurements, as an {\sf IQP} circuit.
{\sf IQP} is defined as a sampling problem 
from the {\sf IQP} circuit,
whose probability distribution is given by
\begin{eqnarray*}
P_{IQP} ( \{ s_i \} | \{ \theta _j\}, \{ S_j \} )
\equiv \left| \bigotimes _{i=1}^{n} \langle +_{s_i} |  \prod_{j} D(\theta _j,S_j) |+\rangle ^{\otimes n}  \right|^2,
\end{eqnarray*}
where $s_i \in \{0, 1\}$ is the measurement outcome 
and $|+_{s_i} \rangle= Z^{s_i} |+\rangle$.
\end{defi}

For each commuting circuit,
we can naturally 
define a bipartite graph $G=(V_A \cup U_B, E)$,
where $V_A$ and $U_B$ are disjoint sets of vertices,
and every edge $\in E$ connects a vertex in $V_A$ 
with another in $U_B$.
Each vertex $v_i \in V_A$ is associated with 
the $i$th input qubit of the {\sf IQP} circuit, and hence $|V_A|=n$. 
Each vertex $u_j \in U_B$ is associated 
with the $j$th commuting gate $D(\theta _j,S_j)$,
and hence $|U_B|= \textrm{poly($n$)}$.
The set of edge $E$ is defined as
$E:=\{ (u_j, v_i) | u_j \in U_B, i  \in S_j  \}$,
that is, the set $S_j$ specifies the vertices $v_i$
that are connected with the vertex $u_j$.
For a given weighted bipartite graph $G=(V_A \cup U_B, E,\{ \theta _j \})$,
where a weight $\theta _j$ is defined on each vertex $u_j \in U_B$,
we can define an {\sf IQP} circuit.

By using Definition~\ref{graphstate} and Remark~\ref{remote-Z},
{\sf IQP} can be rewritten as 
MBQC
on a graph state $|G\rangle$ associated with the graph 
$G=(V_A \cup U_B, E)$. 
In this case, the set $\mathcal{N}_{u_j}$ of vertices 
corresponds to $S_j$.
More precisely,
for a given bipartite graph state $G=(V_A\cup U_B,E)$
and weights $\{ \theta _j\}$,
measurement-based 
{\sf IQP} ({\sf MBIQP}) is defined as follows:
\begin{defi}[MBIQP by Hoban et al.~\cite{Hoban}]
{\sf MBIQP} is defined as a sampling problem according to the probability distribution
\begin{eqnarray*}
P_{MBIQP}(\{m_{v_i}\}, \{ m_{u_j} \} | \{ \theta _{j} \}, G)
\equiv  \left|\bigotimes _{v_i \in V_A} \langle + _{m_{v_i}}|
\bigotimes _{u_j \in U_B} \langle \theta _{j,m_{u_j}} |  | G\rangle \right|^2,
\end{eqnarray*}
where $m_{v_i} \in \{0,1\}$, $m_{u_j} \in \{0,1\}$
and $| \theta _{j,m_{u_j}} \rangle\equiv X^{m_{u_j} } (e^{-i \theta _{j}} |+ _0\rangle + e^{i \theta _{j}}|+_1\rangle)/\sqrt{2}$.
\end{defi}
The bit strings $\{ m_{v_i}\}$ and $\{ m_{u_j}\}$
correspond to the measurement outcomes
on the qubits belonging to $V_A$ and $U_B$, respectively.
We should note that 
there is no temporal order in the measurements
since there is no feedforward of the measurement angles in {\sf MBIQP}.

Then we can prove {\sf MBIQP}={\sf IQP}.
\begin{remk}[{\sf MBIQP} = {\sf IQP} by Hoban et al.~\cite{Hoban}]
{\sf MBIQP} and {\sf IQP} are equivalent in the sense that
if one sampler exists, another sampler can be simulated.
\end{remk}
{\it Proof:} 
Since a stabilizer operator of the graph 
state is given by $K_{u_j} = X_{u_j} \prod _{v_i \in \mathcal{N}_{u_j}} Z_{v_i} $,
$K_{u_j} | G\rangle =| G\rangle$ for each vertex $u_j \in U_B$.
By using this equality,
we obtain
\begin{eqnarray}
P_{MBIQP}(\{m_{v_i}\}, \{ m_{u_j} \} | \{ \theta _{j} \},G)
&=&  \left| \bigotimes _{v_i \in V_A} \langle +_{m_{v_i}} |
\bigotimes _{u_j \in U_B} \langle \theta _{j,m_{u_j}} |    \left( \prod _{u_j \in U_B} K_{u_j}^{m_{u_j}}\right) | G\rangle \right|^2
\nonumber \\
&=&
 \left| \bigotimes _{v_i \in V_A} \langle + _{m_{v_i}} |
\bigotimes _{u_j \in U_B} \langle \theta _{{j},0} |  \left[
\prod _{u_j \in U_B} \left (\prod _{v_i \in \mathcal{N}_{u_j}} Z_{v_i} \right)^{m_{u_j}} \right]  | G\rangle \right|^2
\nonumber \\
&=&
2^{- |U_B|}P_{IQP} (\{  s_{i} \} | \{ \theta _{j} \} , \{S_j \})
\label{MBIQP}
\end{eqnarray}
where $m_{v_i}$ and $s _{i}$ are related via
\begin{eqnarray*}
 s_{i} \equiv m_{v_i} \oplus 
\left( \bigoplus _{u_j \in \mathcal{N}_{v_i}} m_{u_j} \right).
\end{eqnarray*}
In the above, we used the facts that each measurement outcome $\{ m_{u_j} \}$ 
is randomly distributed with probability $1/2$,
and the projection $\langle \theta _{j,0}|$ results in the commuting gate $D(\theta _j,S_j)$ (see Remark~\ref{remote-Z}).
The above equality means that,
regardless of the measurement outcomes $\{ m_{v_i} \}$ and $\{ m_{u_j}\}$,
we can simulate {\sf IQP} by using {\sf MBIQP}.

On the other hand,
by using a random bit string $\{m_{u_j}\}$ with an equal probability 1/2 for each bit and  $\{  s_{i}\}$ sampled from the {\sf IQP} circuit,
we obtain $\{ m_{v_i} \equiv s_{i} \oplus _{u_j \in \mathcal{N}_{v_i}} m_{u_j} \}$
and $\{m_{u_j}\}$,
which is equivalent to the output of {\sf MBIQP}.
\hfill $\square$

As mentioned previously,
there is no feedforward 
for the measurement angles in {\sf MBIQP},
and hence the measurements can be done simultaneously.
This means that {\sf MBIQP} cannot perform universal quantum computation
in {\sf MBIQP} unless 
constant depth circuits can simulate universal quantum computation. 
However, if postselection is allowed,
we can choose the measurement outcomes
in such a way that no byproduct operator is applied.
Thus, with an appropriately chosen
graph structure and weights, 
we can simulate universal quantum computation with the commuting circuits under postselection.
This means that {\sf MBIQP}
with an appropriate graph state and weights (measurement angles)
is universal-under-postselection, and hence post-{\sf MBIQP} = post-{\sf BQP}.
On the other hand, 
Aaronson showed that post-{\sf BQP} = {\sf PP}~\cite{postBQP}.
Accordingly, post-{\sf IQP}=post-{\sf MBIQP} = {\sf PP}.

Here postselected class, post-$A$, is defined 
as a class of decision problems solvable by using 
a computational model 
associated with $A$
(e.g. instantaneous polynomial-time quantum computation for ${\sf IQP}$,
universal quantum computation for ${\sf BQP}$, and
and polynomial-time classical randomized computation for ${\sf BPP}$) 
with a bounded error under postselection.
More precisely,
a language $L$ is in the class post-$A$ 
iff there exists a uniform family $\{ C_w \}$ of circuits of a computational model 
associated with $A$,
where a single line output register $\mathcal{O}_w$ 
(for the $L$-membership decision problem) and a (generally O({\rm poly}(n) )-line) postselection register $\mathcal{P}_w$ are specified
such that 
\begin{enumerate}[(i)]
\item if $w \in L$ then ${\rm Prob}(\mathcal{O}_w =1 | \mathcal{P}_{w} =00...0) \geq 1/2 + \delta$,
\item if $w \notin L$ then ${\rm Prob}(\mathcal{O}_w =1 | \mathcal{P}_{w} =00...0) \geq 1/2 - \delta$,
\end{enumerate}
with a constant $0<\delta<1/2$. 

\begin{figure}[t]
\begin{center}
\includegraphics[width=0.8\textwidth]{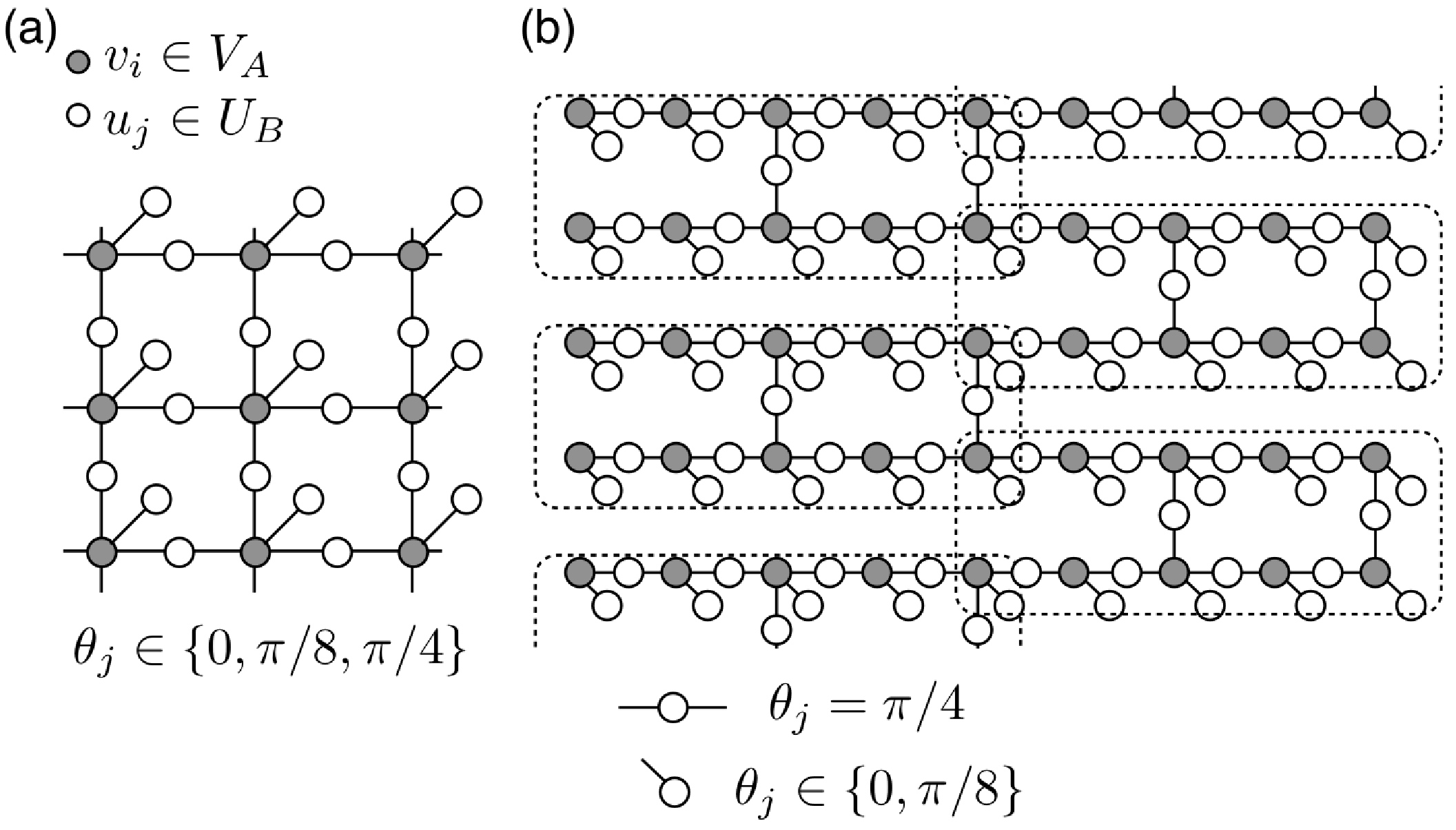}
\end{center}
\caption{(a) An example of a planar bipartite graph and weights for universal {\sf IQP},
where $\theta _j \in \{0, \pi /8, \pi/4\}$. (2) Another example of a planar bipartite graph 
and weights, where $\theta _j=\pi /4$
for all $j$ with $|S_j|=2$ (corresponding to two-qubit commuting gates) and $\theta _j \in \{ 0,\pi/8\}$ for all $j$ with $|S_j|=1$ (corresponding to single-qubit rotations).
The associated graph state is a decollated version of the brickwork state
utilized in blind quantum computation~\cite{Blind,MoriBlind}.
Each dotted square indicates a unit cell of the brickwork state.
The brickwork state allows us to perform universal quantum computation
with measurements only in $\{ |\pm \rangle\}$ and $\{e^{ i (\pi /8) Z}|\pm \rangle\}$ bases.
}
\label{fig0-1}
\end{figure}

In order to simulate post-{\sf BQP},
it is sufficient to consider post-{\sf IQP} or post-{\sf MBIQP}
associated with planar bipartite graphs $G=(V_A\cup U_B, E)$
with $|S_j|\leq 2$ and $\theta _j = \pi /8$ for all $j$~\cite{IQP}.
(As shown in Sec.~\ref{sec5}, 
we can obtain the same result not only for $\theta _j =\pi/8$ but also for almost all angles $\theta _j$.)
In this case, each instance is encoded into a structure of a graph. 
In another encoding, we can fix the structure of the graph
but choose each angle $\theta _j$ from $\{ \pi /4, \pi /8, 0\}$.
Specifically, $\theta _j=0$ corresponds to a deletion of 
vertex $u_j$ from the graph (see Remark~\ref{Zmeasurement}). 
$\theta _j = \pi /4$ and $\pi /8$ correspond to 
Clifford and non-Clifford gates, respectively.
Examples of graphs and weights of {\sf MBIQP} that 
are universal-under-postselection are presented in Fig.~\ref{fig0-1} (a) and (b).

In Ref.~\cite{IQP}, Bremner, Jozsa, and Shepherd showed that
if {\sf IQP} is weakly simulatable 
by using a classical randomized algorithm with a multiplicative approximation error $1<c<1/\sqrt{2}$:
\begin{eqnarray*}
\frac{1}{c} P_{IQP} \leq P^{\rm ap}_{IQP} \leq c P_{IQP},
\end{eqnarray*}
then the {\sf PH}
collapses to the third level. 
The {\sf PH}
is a natural way of classifying the complexity of problems
(languages) beyond {\sf NP} (nondeterministic polynomial-time computation).
The level-$k$ class $\Delta _k$
of the hierarchy is defined recursively 
by $\Delta _{k+1} = \textrm{\sf P}^{\textrm{N}\Delta _k}$.
Then the {\sf PH} is defined as
the union ${\sf PH}\equiv \cup_k \Delta _k$ of them.
Here {\sf A}$^\textrm{\sf B}$ indicates computation {\sf A} with 
an oracle for {\sf B} and ${\sf ``N"A}$ means the nondeterministic version of {\sf A}.
{\sf NP}={\sf P} implies a collapse of the {\sf HP} at the first level,
that is, the {\sf PH} collapses completely.
The collapse of the {\sf PH} to the third level is not as unlikely as
{\sf NP}={\sf P} but still thought to be highly implausible.

\begin{remk}[Hardness of {\sf IQP} by Bremner et al.~\cite{IQP}]
\label{IQP}
If {\sf IQP}
is weakly simulatable by a classical polynomial time randomized algorithm within multiplicative error 
$1\leq c \leq \sqrt{2}$, {\sf PP} $=$ post-{\sf BPP},
resulting in a collapse of the {\sf PH} to the third level.
\end{remk}
{\it Proof}: (See also Ref.~\cite{IQP}.)
Let $L$ be a language
decided by post-{\sf IQP} with a bounded error $0<\delta<1/2$,
that is,
\begin{eqnarray}
\textrm{if $w \in L$, }
P_{IQP}( \mathcal{O}_w= 1 | \mathcal{P}_w =00...0 ) \geq 1/2 +\delta,
\label{decision1}
\\
\textrm{if $w \notin L$, }
P_{IQP}( \mathcal{O}_w= 1 | \mathcal{P}_w =00...0) \leq 1/2 -\delta,
\label{decision2}
\end{eqnarray}
with a constant $0 < \delta < 1/2$.
Suppose we have 
a classical polynomial-time randomized algorithm that 
weakly simulates {\sf IQP}, i.e., a sampling 
according to the probability distribution $P_{IQP}(\mathcal{O}_w=x, \mathcal{P}_w=y)$ 
with a multiplicative error $1<c<\sqrt{2}$.
Under postselection,
we can simulate post-IQP,
a sampling according to the probability distribution
\begin{eqnarray*}
P^{\rm ap}_{IQP}( \mathcal{Q}_w=x | \mathcal{P}_w =00..0)=\frac{P^{\rm ap}_{IQP}(\mathcal{O}_w=x,\mathcal{P}_w=00...0)}{P^{\rm ap}_{IQP}(\mathcal{P}_w=00...0) }.
\end{eqnarray*}
The multiplicative error 
for the conditional probability $P^{\rm ap}_{IQP}( \mathcal{Q}_w=x | \mathcal{P}_w =00..0)$ is 
bounded by $c^2$:
\begin{eqnarray*}
\frac{1}{c^2} P_{IQP}( \mathcal{Q}_w=x | \mathcal{P}_w =00..0)
\leq P^{\rm ap}_{IQP}( \mathcal{Q}_w=x | \mathcal{P}_w =00..0)
\leq c^2 P_{IQP}( \mathcal{Q}_w=x | \mathcal{P}_w =00..0).
\end{eqnarray*}
Using this and Eqs. (\ref{decision1}) and (\ref{decision2}), 
we obtain 
\begin{eqnarray*}
\textrm{if $w \in L$, }
P^{\rm ap}_{IQP}( \mathcal{Q}_w=1 | \mathcal{P}_w =00..0)\geq 
\frac{1}{c^2} (1/2+\delta) , 
\\
\textrm{if $w \notin L$, }
P^{\rm ap}_{IQP}( \mathcal{Q}_w=1 | \mathcal{P}_w =00..0)\leq 
c^2 (1/2-\delta) . 
\end{eqnarray*}
Thus if both 
$c^{-2} (1/2+\delta ) > 1/2$  
and $c^{2} (1/2-\delta ) <1/2$ are satisfied,
we can construct a classical randomized algorithm that
decides $L$ with bounded error.
In other words, post-{\sf IQP} $\subseteq$ post-{\sf BPP}.
Since post-{\sf IQP} does not 
depend on the level of error $\delta$,
we can choose any value $0<\delta <1/2$.
By using the fact that {\sf IQP} is 
universal-under-postselection, 
we conclude that if $c<\sqrt{2}$, 
{\sf PP} = post-{\sf BQP} =  post-{\sf IQP} $\subseteq$ post-{\sf BPP}.
Apparently, post-{\sf BQP} includes post-{\sf BPP},
and hence {\sf PP} = post-{\sf BPP}.

Due to Toda's theorem~\cite{Toda},
{\sf P} with an oracle
for {\sf PP} includes 
whole classes in the {\sf PH},
i.e., {\sf PH} $\subseteq$ {\sf P}${}^{{\sf PP}}$.
On the other hand, {\sf P} with an oracle for post-{\sf BPP}
is in the third level of the {\sf PH}, i.e, P${}^{\rm post-{\sf BPP}} \subseteq 
\Delta _3$.
Thus {\sf PP} = post-{\sf BPP}
implies a collapse of the {\sf PH} to the third level,
which is highly implausible.
In other words, unless the {\sf PH} collapses to the third level,
there exists no efficient weak classical simulation of {\sf IQP}.
\hfill $\square$

\subsection{Strong simulation and post-{\sf BQP} = {\sf PP} theorem}
Aaronson's theorem, post-{\sf BQP} = {\sf PP}~\cite{postBQP}, is quite useful to obtain
not only quantum complexity results 
combined with the postselection argument by Bremner, Jozsa, and Shepherd~\cite{IQP},
but also to provide ``quantum proofs" of classical complexity results~\cite{QuantumProof}.
For example, in Ref.~\cite{postBQP}, Aaronson 
provided alternative and much simpler proof that {\sf PP} is closed under intersection~\cite{PPclosed}.
Moreover, by using post-{\sf BQP} = {\sf PP}, 
we can show that strong simulation of some computational tasks,
which are as hard as post-{\sf BQP} under postselection,
is \#{\sf P}-hard even in an approximated case with a multiplicative error:

\begin{remk}[Strong simulation and post-{\sf BQP} = {\sf PP}]
\label{StrongpostBQP}
Suppose a (classical or quantum) computation $A$ is universal-under-postselection
and has enough postselection ports,
so that post-$A$ = post-{\sf BQP}.
Then an exact strong simulation of $A$ is as hard as 
an exact strong simulation of the output of universal quantum computer
and hence \#{\sf P}-hard. Moreover, an approximated strong simulation of $A$ 
with a multiplicative error $1<c<\sqrt{2}$ is also \#{\sf P}-hard. Thus if the output 
of $A$ is efficiently strongly simulatable (or equivalently if there is a fully polynomial-time classical approximation scheme for the output distribution of $A$), 
\#{\sf P}-hard problems are solved efficiently, and hence the {\sf PH} collapses completely.    
\end{remk}

{\it Proof:}
Suppose the probability distribution $P_{A}(\mathcal{O}_w=x, \mathcal{P}_w=y)$ of the output of $A$
can be strongly simulated with a multiplicative error $1<c<\sqrt{2}$:
\begin{eqnarray*}
\frac{1}{c} P_{A}(\mathcal{O}_w=x, \mathcal{P}_w=00...0) \leq 
P^{\rm ap}_{A}(\mathcal{O}_w=x, \mathcal{P}_w=00...0) \leq c P_{A}(\mathcal{O}_w=x, \mathcal{P}_w=00...0).
\end{eqnarray*} 
By using this, we can calculate the postselected probability distribution
\begin{eqnarray*}
P^{\rm ap}_{A}(\mathcal{O}_w=x | \mathcal{P}_w=00...0)  
= \frac{ P^{\rm ap}_{A}(\mathcal{O}_w =x, \mathcal{P}_w=00...0)}{\sum _{ x' =0,1} 
P^{\rm ap}_{A}(\mathcal{O}_w=x' , \mathcal{P}_w=00...0)}
\end{eqnarray*} 
with a multiplicative error $1<c^2<2$.
Since post-$A$ $=$ post-{\sf {BQP}} $=$ {\sf PP},
if we can calculate $P^{\rm ap}_{A}(\mathcal{O}_w=x | \mathcal{P}_w=00...0)$ efficiently
with a multiplicative error $c^2<2$,
it is sufficient to decide a complete problem in {\sf PP}.
Since ${\sf P}^{\sf PP} = {\sf P}^{\sf \#P}$,
the multiplicative approximation is enough to find a solution of $\#P$-complete
problem and hence $\#P$-hard.
Moreover, the multiplicative approximation 
results in an entire collapse of the {\sf PH}.

The above remark indicates that if 
a function $f(x)$ of interest is given as a probability distribution 
of some quantum task that is universal-under-postselection,
then computation of $f(x)$ is \#{\sf P}-hard 
even in the approximated case with a multiplicative error.
This argument has been utilized by Kuperberg to show \#{\sf P}-hardness of approximating 
the Jones polynomial with a multiplicative error~\cite{Kuperberg}.
In Ref.~\cite{AaronsonLinear},
Aaronson provided an alternative proof of \#{\sf P}-hardness of calculating the permanent~\cite{Valiant}
based on the above argument and the KLM scheme~\cite{KLM}.
We will also utilize it to provide the \#{\sf P}-hardness of a multiplicative approximation
of Ising partition functions with an imaginary parameter region,
in Sec.~\ref{sec5}.
Moreover, Remark~\ref{StrongpostBQP} also implies 
that there is a good chance for a quantum computer
in an approximation a function $f(x)$ with an additive error 
under an appropriate normalization through 
the Hadamard test~\cite{AharonovJones,AharonovJonesHard,AharonovTutte}.

\subsection{Related works}
As a final part of the preliminary section,
we review related works on computational complexity of commuting quantum circuits 
and Ising partition functions.

In Ref.~\cite{NiNestCommuting},
they have investigated rather general commuting quantum circuits 
of $d$-level (qudit) systems. Not only the diagonal gates in the computational basis, 
but also general commuting gates are considered.
Specifically they showed that 
a single qudit output (or at most polylogarithmic number of qudits)
of 2-local commuting quantum circuits is strongly simulatable 
with an exponential accuracy.
Moreover, a single qudit output of 3-local commuting quantum circuits 
cannot be strongly simulated, 
unless every problem in \#{\sf P} has a polynomial-time classical algorithm. 
The former result and intractability of {\sf IQP} with two-local commuting gates
imply that a polynomial size of the output is essential 
for commuting quantum circuits to be hard for a weak classical simulation.

In Ref.~\cite{NakataDiag},
it has been shown that an approximated random state, $t$-design, 
can be generated by diagonal (i.e., commuting) quantum circuits~\cite{NakataDiag0,NakataDiag1}
(see also a review~\cite{NakataDiag}).
Since random states are shown to be useful in various quantum information tasks~\cite{RandomAlg0,RandomAlg1,RandomAlg2},
they are one of the most important applications of commuting quantum circuits.

For the ferromagnetic Ising models with a constant magnetic field on arbitrary graphs, 
there exists 
a fully polynomial-time randomized approximation scheme (FPRAS)~\cite{Jerrum1993},
which approximates
the partition function $Z_{\rm Ising}$ of the size $n$
with a multiplicative error $c=1+\epsilon$ in a ${\rm poly}(n,1/\epsilon)$ time. 
However, under the random magnetic fields, approximation of ferromagnetic Ising 
partition functions below a certain critical temperature equivalent, under an approximation-preserving reduction,
to \#{\sf BIS}, which is a counting problem of 
the number of independent sets of a bipartite graph~\cite{Goldberg2007}.
The counting problem \#{\sf BIS} is conjectured to lie in-between FPRAS and \#{\sf SAT} under an approximation-preserving reduction.
Here \#{\sf SAT} indicates a counting problem of the number of satisfying configurations,
and does not 
have an efficient (polynomial) multiplicative approximation unless {\sf NP}={\sf RP}~\cite{Zuckerman}.
Moreover, it has been shown that a multiplicative approximation 
of antiferromagnetic Ising partition functions (below a certain threshold temperature)
on $d$-regular graphs ($d \geq 3$) are {\sf NP}-hard~\cite{Sly}.
A comprehensive classification of complexity of multiplicative approximation of
complex-valued Ising partition functions has been provided in Ref.~\cite{GG}.

In Ref.~\cite{Lidar97}, a quantum algorithm to prepare quantum states
encoding the thermal states of Ising models 
has been proposed for a restricted type of lattice structures.
In Ref.~\cite{Lidar04}, it has been shown that
calculations of partition functions of $\pm J$ random-bond Ising models
are equivalent to quadratically signed weight enumerators,
with an oracle for which 
classical probabilistic computation is polynomially equivalent to quantum computation~\cite{KnillQSWE}.
Based on this mapping,
certain quantum circuits corresponding to
Ising models on planar lattices without magnetic fields
have been shown to be efficiently simulatable by a classical computer in the strong sense~\cite{Lidar10}.

Quantum algorithms to approximate the Ising partition functions 
in a complex parameter region have been studied so far using 
a transfer matrix method~\cite{Nest_circuit,Cuevas11}, an overlap mapping ~\cite{VandenNest07,completeness,FujiiStat,Matsuo14},
and a path integral method~\cite{Iblisdir}.
Specifically, certain sets of instances are shown to be BQP-complete,
which means that such algorithms can actually do a nontrivial task,
which would be intractable on a classical computer.
In Ref.~\cite{Iblisdir},
a quantum algorithm for an additive approximation of real Ising partition functions 
on square lattices
has been proposed by using an analytic continuation (see also a Fourier sampling scheme 
for spin models for estimating free energy~\cite{Yamamoto03}).
In Ref.~\cite{Matsuo14},
another quantum algorithm for an additive approximation of 
square-lattice Ising partition functions with 
completely general parameters including real physical ones
has been constructed 
based on a linear operator simulation by a unitary circuit
with ancilla qubits (see also a linear operator simulation 
for an additive approximation of Tutte polynomials~\cite{AharonovTutte}).
Specifically, in this case, the achievable approximation scale
was also calculated explicitly.
The Ising partition functions on square lattices
with magnetic fields
are know to be universal in the sense that
the partition function of any other classical spin model can be mapped into 
an Ising partition function by choosing a certain parameter~\cite{completeness}.
Thus the above quantum algorithm allows approximation of 
an arbitrary classical spin partition function
with a certain approximation scale.

\section{Bridging {\sf IQP} and Ising partition functions}
\label{sec3}
In this section,
we establish a bridge between 
{\sf IQP} and Ising partition functions.
We will first show that 
the joint probability distribution of the output of an {\sf IQP} circuit
associated with a graph $G$
is given by normalized squared norm of the partition function
of the Ising model defined by the graph $G$.
This is shown 
first by mapping {\sf IQP} into {\sf MBIQP}
and then by using the overlapping map~\cite{VandenNest07},
which relates
the Ising partition functions with
an inner product between a product state 
and the graph state $|G\rangle$.
However, this is not sufficient for our purpose.
Since there are exponentially many instances of the measurement outcomes,
a straightforward sampling using the joint probability distributions 
does not work efficiently.
Instead, we simulate {\sf IQP}
in a recursive way according to the conditional distribution 
on the previous measurement outcomes
by using the Bayes theorem.
To this end, we need 
the marginal distributions 
with respect to the measured qubits.
If the marginal distribution can be calculated 
efficiently, the recursive method succeeds to 
simulate a sampling according to the joint probability distribution of {\sf IQP} efficiently.
In this section, we will also establish a relationship 
between the marginal distribution 
with respect to a set $M$ of the measured qubits
and the Ising partition function
defined on another graph $\tilde G_M$,
which is systematically
constructed from the graph $G$ and the set $M$.

\subsection{Joint probability distribution}
We define an Ising model,
which may include multibody interactions,
according to the bipartite graph $G=(V_A\cup U_B,E)$ and weights $\{\theta _j\}$.
The Ising model consists of
the sites associated with the vertices $v_i \in V_A$
and
multibody interactions represented by the vertices $u_j \in U_B$.
The spins engaged in the $j$th interaction
and its coupling constant
are given by $\mathcal{N}_{u_j}$ (or equivalently $S_j$)
and $\theta _j$, respectively.
\begin{defi}[Multibody Ising Model with random $i\pi/2$ magnetic fields]
For a given bipartite graph $G=(V_A\cup U_B, E)$
and weights $\{\theta _j\}$ defined on the vertices in $U_B$,
a Hamiltonian of an Ising model with random $i \pi /2$ magnetic fields
is defined by
\begin{eqnarray}
H(\{s_{i}\}, \{ \theta _{j} \}, G) 
\equiv -  
\sum_{v_i \in V_A}  i\pi s_{i} \frac{1-\sigma _{v_i}}{2}
- \sum _{u_j \in U_B} i  \theta _{j}\left(  \prod _{v_i  \in \mathcal{N}_{u_j} } \sigma _{v_i} \right),
\label{Isingmodel}
\end{eqnarray}
where $\sigma _{v_i} \in \{ +1,-1\}$ is an Ising variable
defined on each vertex $v_i \in V_A$.
The partition function 
of the Ising model is defined by
\begin{eqnarray*}
\mathcal{Z}( \{s_{v_i}\}, \{ \theta _{j} \}, G)
= \sum _{\{ \sigma _{v_i} \}} e^{ - H(\{s_{i}\}, \{ \theta _{j} \}, G)},
\end{eqnarray*}
where $\sum _{\{ \sigma _{v_i} \}}$ means 
the summation over all configuration $\{\sigma _{v_i} \}$.
\end{defi}
We should note that, in addition to 
the interactions defined by the graph and weights,
random $i \pi /2$ magnetic fields 
are also introduced according to the bit string $\{ s_{v_i}\}$.
This corresponds to the measurement outcome of ${\sf IQP}$
as seen below. Furthermore, in Sec.~\ref{sec4},
these random $i\pi/2$ magnetic fields will 
be successfully removed for a certain class of Ising models
by renormalizing them into the coupling constants 
$\{ \theta _j \}$.

The probability distribution of {\sf IQP}
associated with $G=(V_A\cup U_B,E)$ and weights $\{ \theta _j\}$
is now shown to be equivalent to the normalized squared norm of 
the partition function of Ising model defined by the graph $G$
and weights $\{ \theta _j\}$
as follows:
\begin{theo}[{\sf IQP} and Ising partition functions]
\label{Main1}
{\sf IQP} associated with the graph $G=(V_A\cup U_B,E)$ 
and weights $\{\theta _j\}$ is equivalent to the sampling 
problem according to the normalized squared norm of an
Ising partition function defined by the graph $G$ and weights $\{ \theta _j\}$:
\begin{eqnarray*}
P_{IQP}(\{s_i \} | \{ \theta _{j} \}, \{ S_j \})
&=&2^{ |U_B|} P_{MBIQP} 
(\{m_{v_i}\}, \{ m_{u_j}\} | \{ \theta _{j} \}, G)
\\
&=&2^{-2|V_A|}
\left|\mathcal{Z}(\{ s_{i}\}, \{ \theta _{j} \}, G)\right|^2.
\end{eqnarray*}
\end{theo}
{\it Proof:}
We reformulate 
the left hand side of Eq. (\ref{MBIQP}) using the overlap mapping 
developed by Van den Nest, D{\"u}r, and Briegel~\cite{completeness,FujiiStat}:
\begin{eqnarray}
&& P_{IQP}(\{s_i\} | \{ \theta _{j} \}, \{S_j \})
\nonumber \\
&=&2^{|U_B|}P_{MBIQP}(\{m_{v_i}\}, \{ m_{u_j} \} | \{ \theta _{j} \},G)
\nonumber \\
&=&  2^{|U_B|}
\left| 
\left(
 \bigotimes _{v_i \in V_A} 
\langle + _{s_i} | \right)
\left(
\bigotimes _{u_j \in U_B} 
 \langle  \theta _{j,0} |H \right)
\prod _{u_j \in U_B} H_{u_j}| G \rangle \right|^2
\nonumber \\
&=& 2^{|U_B|}
\left|
  \left(\bigotimes _{v_i \in V_A}  
\frac{\langle 0| + e^{i s_i  \pi  } \langle 1| }{\sqrt{2}} 
  \right)
  \left(
\bigotimes _{u_j \in U_B} 
\frac{\langle 0 | e^{  i \theta _{j}} +\langle 1| e^{- i \theta _{j}}}{\sqrt{2}} 
\right) \left(
2^{-|V_A|/2}\sum _{\{ \bar \sigma _{v_i}  \} }  |\{ \bar \sigma _{v_i} \} \rangle 
\bigotimes_{u_j\in U_B}\left| \bigoplus _{ v_i \in N_{u_j} } \bar \sigma _{v_i} \right\rangle \right)   \right|^2
\nonumber \\
&=& 2^{|U_B|}
\left|
2^{-|U_B|/2-|V_A|} 
\sum _{\{ \bar \sigma _{v_i} \} }   
\exp\left[\sum_{v_i \in V_A}  i\pi s_{i} \bar \sigma _{v_i} \right]
\exp \left[ \sum _{u_j \in U_B }  -i \left[2 \theta _{j} \left(  \bigoplus _{ v_i \in N_{u_j} } \bar \sigma _{v_i} \right) -  \theta _j \right]   \right] \right|^2
\nonumber \\
&=& 
2^{-2|V_A| }
\left|\sum _{ \{ \sigma _i \}} e^{-  H(\{s_{i}\}, \{ \theta _{j} \},G) } \right|^2
\nonumber \\
&= &
2^{-2|V_A| }
\left|\mathcal{Z}(\{s_{i}\}, \{ \theta _{j} \},G)\right|^2,
\label{Mapping}
\end{eqnarray}
where we define a binary variable $\bar \sigma _{v_i} \equiv (1-\sigma _{v_i})/2$,
and $\sum _{\bar \sigma _{v_i}}$ indicates
a summation over all binary strings.
From the second to the third lines,
we used the fact that 
\begin{eqnarray*}
|G\rangle &=& \left( \prod _{u_j \in U_B}  \prod _{v_i \in \mathcal{N}_{u_j} }
\Lambda _{v_i,u_j}(Z) \right) |+\rangle ^{\otimes |V_A|} |+\rangle ^{\otimes |U_B|}
\\
&=&
\left( \prod _{u_j \in U_B} H_{u_j}\right) \left( \prod _{u_j \in U_B}  \prod _{v_i \in \mathcal{N}_{u_j} }
\Lambda _{v_i,u_j}(X) \right)
 \sum _{\{ \bar \sigma _{v_i}\}} | \{ \bar \sigma _{v_i}\} \rangle 
| 0 \rangle ^{ \otimes |U_B|}
\\
&=&
\left( \prod _{u_j \in U_B} H_{u_j}\right) 
 \sum _{\{ \bar \sigma _{v_i}\}} | \{ \bar \sigma _{v_i}\} \rangle 
\bigotimes _{u_j \in U_B} | \oplus _{v_i \in \mathcal{N}_{u_j} } \bar \sigma _{v_i }  \rangle.
\end{eqnarray*}
\hfill $\square$
\\

Equation (\ref{Mapping}) shows that
{\sf IQP} is equivalent to 
the sampling problem according to the probabilities proportional to the squared norm of the partition functions of an Ising model with imaginary coupling constants.
Note that the measurement outcome $\{ s_i \}$
correspond to the random $i \pi /2$ magnetic fields.

The present sampling problem is not
related directly to
what is well studied in the fields of statistical physics,
such as the Metropolis sampling according to the Boltzmann distribution.
However, as we will see below,
the relation between {\sf IQP} and Ising partition functions
leads us to several interesting results
about complexity of {\sf IQP},
since calculation of the Ising partition functions 
are well studied in both fields of statistical physics and 
computer science.
It was shown
in Ref.~\cite{Barahona} that 
exact calculation of partition functions of two-body Ising models 
with magnetic fields even on the planar graphs is NP-hard.
Furthermore, in general,
exact calculation of partition functions of 
two-body Ising models with magnetic fields
is \#{\sf P}-hard~\cite{Jerrum93}.
No polynomial-time approximation scheme with 
multiplicative error exists unless {\sf NP}={\sf RP}.
While {\sf IQP} does not provide the exact values 
of the partition functions,
it is surprising that the sampling
according to the partition functions
of many-body Ising models 
$H(\{s_{v_b}\}, \{ \theta _{v_a} \},G)$
with imaginary coupling constants,
can be done in {\sf IQP}, which
consists only of commuting gates
and seems much weaker than {\sf BQP}.

Only in the limited cases, the partition function of 
an Ising model can be calculated efficiently.
Such an example is two-body Ising models
on the 2D planar lattices without magnetic fields.
In the next section,
we show that certain classes of {\sf IQP}
are classically simulatable, at least in the weak sense, by using the fact that
the associated Ising models are exactly solvable.
To this end, we need not only 
the joint distribution of the output of {\sf IQP}
circuits but also the marginal distributions 
with respect to measured qubits,
in order to simulate the sampling problem recursively.

\subsection{Marginal distribution}
\begin{figure}[t]
\begin{center}
\includegraphics[width=0.9\textwidth]{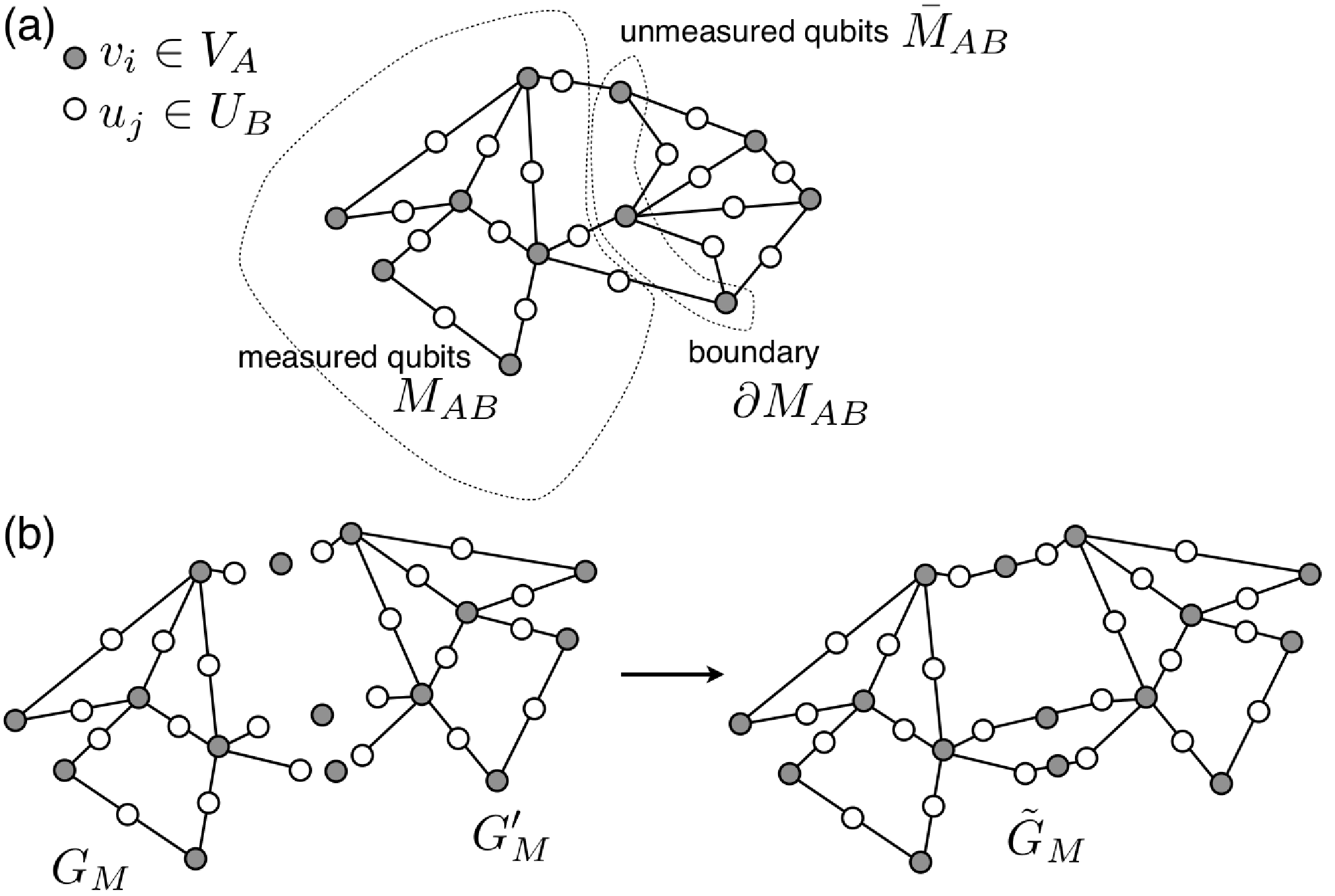}
\end{center}
\caption{(a) The graph state $|G\rangle$ associated with the graph $G$.
The gray and white circles indicate qubits associated with $v_i \in V_A$ and $u_j \in U_B$,
respectively. (b) The subgraph state $|G_M\rangle$ and its copy $|G'_M\rangle$ 
are merged via the qubits $|+\rangle ^{ |\partial M_{AB}|}$ located on the boundary.
The merged graph is denoted by $\tilde G_M$.} 
\label{fig3}
\end{figure}

Even if we can calculate
the probability distribution
$P_{IQP}(\{ s_i \} | \{ \theta _j \} ,\{S_j\})$
efficiently,
it does not directly mean that the corresponding {\sf IQP} 
is classically simulatable,
since there are exponentially many 
varieties of the measurement outcomes $\{s_i \}$.
An efficient weak classical simulation of 
{\sf IQP} requires the marginal distribution
with respect to measured qubits, by which 
we can simulate {\sf IQP} recursively.
In the following we 
will establish a mapping 
between the marginal distribution with respect to the 
set $M$ of measured qubits and 
the partition function of an Ising model 
defined on a merged graph $\tilde G_{M}$.
The merged graph $\tilde G_{M}$ constructed by merging
a subgraph $G_M$
corresponding to the measured part of the graph $G$
and its copy $G'_M$ (see Fig.~\ref{fig3}).
(The detailed definition of the subgraph
$G_M$ and the merged graph $\tilde G_M$ are given in
the proof of the following theorem.)

\begin{theo}[Marginal distribution of {\sf IQP}]
\label{Main2}
Let $M \subset \{ 1,2,..., n\}$ and $\bar M \subset \{ 1,2,...,n\}$ be sets of the measured and unmeasured qubits, 
respectively (and hence $M \cup \bar M = \{ 1,2,...,n\}$ and $M \cup \bar {M} = \emptyset$).
A marginal distribution 
with respect to the set $M $
\begin{eqnarray*}
P_{IQP}(\{ s_ i\}_{i \in M}| \{ \theta _j \} , \{S_j\},  M ) \equiv \sum _{ \{ s_i \}_{ i \in  \bar M} } 
P_{IQP}(\{ s_i \} | \{ \theta _j \} , \{ S_j\} )
\end{eqnarray*}
is related to
the Ising partition function 
defined by the merged graph ${\tilde G_M}$ and weights $\{\theta _j\} \cup \{- \theta  _j \}$.
\end{theo}
{\it Proof:}
In order to prove this,
we consider the corresponding {\sf MBIQP}.
However, it is just for a proof,
and hence we do not need to simulate {\sf MBIQP}
in classical simulation as seen later.
Thus without loss of generality, we 
can assume that the measurement outcome is subject to
$m_{u_j}=0$ for all $u_j \in U_B$.

Based on the sets $M$ and $\bar M$,
the sets of measured and unmeasured qubits in 
$V_A$ is defined as $M_A$ and $\bar M_A$, i.e., $M_A \cup \bar M_A =V_A$.
We define a subgraph $G_M (M_A \cup M_B, E_M)$,
where $M_B \subset U_B$ is a set of vertices 
that are connected with any vertices in $M_A$,
i.e., $M_B=\{ u_j\in U_B | (u_j,v_i) \in E, v_i \in M_A \}$.
$E_M$ is a set of edges whose two incident vertices both belong to
$M_A\cup M_B$.
We denote $M_A \cup M_B$ simply by $M_{AB}$
and $(V_A \cup U_B)\backslash M_{AB}$ by $\bar M_{AB}$ (see Fig.~\ref{fig3} (a)).

The marginal distribution can be 
written as
measurements on the reduced density matrix on
the qubits $M_{AB}$:
\begin{eqnarray*}
P_{IQP}(\{ s_{i}\}_{i \in  M}| \{ \theta _{j} \} , \{ S_j \} ,   M )
&=&
\langle \Theta |
{\rm Tr}_{\bar M_{AB}}\bigl[|G\rangle \langle G| \bigr] | \Theta \rangle,
\end{eqnarray*} 
where $| \Theta \rangle 
\equiv \bigotimes _{v_i \in M_A} |+_{s_{i}} \rangle 
\bigotimes _{u_j \in M_B} |\theta _{j,0}\rangle $,
and ${\rm Tr}_{\bar M_{AB}}$ indicates 
the partial trace with respect to the unmeasured qubits $\bar M_{AB}$.

We define a subset $\partial M_{AB} \subset \bar M_{AB}$
as a set of vertices connected with any vertices in $M_{AB}$,
i.e.,
 $\partial M_{AB} = \{ v_i \in \bar M_{A}
|(v_i, u_j) \in E, u_j \in M_{B} \}$
(note that $\partial M_{AB} \subset \bar M_A$).
We refer to the qubits associated with 
the vertices in $\partial M_{AB}$ as
the boundary qubits,
since they are the boundary of the 
measured and unmeasured qubits in the graph state
as shown in Fig.~\ref{fig3} (a).

For the graph state $|G\rangle$,
the tracing out with respect to the unmeasured qubits $\bar M_{AB}$
can be equivalently done by $Z$ basis measurements
on the boundary qubits and forgetting about the 
measurement outcomes. This is because,
$Z$-basis measurements on the boundary qubits separate 
the measured and unmeasured qubits (see Remark~\ref{Zmeasurement}),
and hence the tracing out of the qubits in $\bar M_{AB} \backslash \partial M^{AB}$
does not have any effect on the measured qubits $M_{AB}$.
From this observation we obtain
\begin{eqnarray*}
{\rm Tr}_{\bar M_{AB}}\bigl[|G\rangle \langle G| \bigr]
=2^{ -|\partial M_{AB}|} \sum _{\{m_{v_i} \}_{\partial M_{AB} }} 
\left(\prod _{v_i \in \partial M_{AB} } B({v_i})^{m_{v_i}} \right) |{G_M} \rangle \langle {G_M} |\left(\prod _{v_i \in \partial M_{AB} } B({v_i})^{m_{v_i}} \right) 
\end{eqnarray*}
where $\{  m_{v_i} \}_{\partial M_{AB} }$
is the set of the measurement outcomes 
on the boundary qubits, and 
we define a byproduct operator
$B(v_i) = \prod _{u_j \in \mathcal{N}_{v_i} \cap M_{AB}} Z_{u_j}$
(see Remark~\ref{Zmeasurement}).

Let us consider a merged graph $\tilde G_M$ 
that is constructed from the graph $G_M$
and its copy $G'_M$, and the boundary $\partial M_{AB}$.
Two copies of graph states, $|G_M\rangle$ and $|G'_M\rangle$, 
are merged via $|+\rangle^{\otimes |\partial M_{AB}|}$ 
as shown in Fig.~\ref{fig3} (b).
The vertices in $\partial M_{AB}$ and 
those in $G_M$ and $G'_M$ are connected
iff there is an edge between them
in the original graph $G$ and its copy $G'$.
The graph state associated with the merged graph $\tilde G_M$
is written as
\begin{eqnarray*}
|\tilde G_M \rangle  = \prod _{v_i \in \partial M_{AB} }
\left( \prod_{ u_j \in \mathcal{N}_{v_i}\cup M_{B} } \Lambda _{v_i,u_j} (Z) 
 \prod_{ u'_j \in \mathcal{N}'_{v_i}\cup M'_{B} } \Lambda _{v_i,u'_j} (Z) \right)
 |G_M\rangle |+\rangle ^{\otimes 
|\partial M_{AB} |} | G'_M \rangle.
\end{eqnarray*}

Let us consider a projection of $|\tilde G_M\rangle$
by $|+\rangle ^{\otimes |\partial M_{AB} |}$:
\begin{eqnarray*}
\langle + | ^{ \otimes | \partial M_{AB} |} |\tilde G_M \rangle  
= 
2^{- |\partial M_{AB}|}
\sum _{\{  m_{v_i} \} _{\partial M_{AB} }}
\left[ \prod _{v_i \in \partial M_{AB}} \left[B(v_i) 
 {B'}(v_i) \right]^{ m_{v_i}}\right] |G_M\rangle  | G'_M \rangle,
\end{eqnarray*}
where $B'(v_i)$ is defined similarly to $B(v_i)$
on the graph state $|G'_M\rangle$.
Let us define 
\begin{eqnarray*}
|\Theta ' \rangle \equiv  
\bigotimes _{v_i \in M_A} |+_{s_{i}} \rangle 
\bigotimes _{u_j \in M_B} |-\theta _{j,0}\rangle ,
\end{eqnarray*}
where we should note that the sign of the angle $\theta _{j,0}$ 
is flipped.
Next we consider 
a projection by $| \Theta \rangle | \Theta ' \rangle$ as follows:
\begin{eqnarray}
&&\langle \Theta | \langle + |^{ \otimes | \partial M_{AB} |} \langle \Theta ' | | \tilde G_M\rangle
\nonumber\\
&=&
2^{- |\partial M_{AB}|}
\sum _{\{  m_{v_i} \} _{\partial M_{AB} }}
\langle \Theta |\left[ \prod _{v_i \in \partial M_{AB}} \left[B(v_i) 
 \right]^{ m_{v_i}}\right] |G_M\rangle  
\langle \Theta ' |  \left[ \prod _{v_i \in \partial M_{AB}} \left[B'(v_i) 
 \right]^{ m_{v_i}}\right] 
 | G'_M \rangle
\nonumber
\\
&=& \langle \Theta | {\rm Tr} _{\bar M_{AB} }\bigl[|G\rangle \langle G| \bigr] 
|\Theta \rangle 
\nonumber
\\
&=& P_{IQP}(\{ s_{i}\}_{i \in M}| \{ \theta _{j} \} , \{S_j \}, M ).
\label{overlap1}
\end{eqnarray}
This indicates that the summation over exponentially many variables for the marginalization
is taken simply in an overlap between the product state and the merged graph state.

On the other hand,
the overlap 
$\langle \Theta | \langle + |^{ \otimes | \partial M_{AB} |} \langle \Theta ' | | \tilde G_M\rangle$ is also 
reformulated as an Ising partition function 
as done in the proof of Theorem~\ref{Main1}.
Specifically, the interaction patterns 
are given by the merged graph $\tilde G_M$.
The coupling strengths are given by 
two copies of $\{ \theta _{j} \}_{u_j \in M_B}$ and $\{ -\theta _{j} \}_{u'_j \in M'_B}$:
\begin{eqnarray}
&&\langle \Theta | \langle + |^{ \otimes | \partial M_{AB} |} \langle \Theta ' | | \tilde G_M\rangle 
\nonumber
\\
&=& 2^{-2|M_A|-|\partial M_{AB}| - |M_B|}
\left| \mathcal{Z} ( \{ s_{i}\} _{M} \cup \{ 0 \}_{ v_i \in \partial M_{AB}} \cup \{s'_i \}_{M'} , \{\theta _{j} \}_{u_j \in M_B} \cup \{ -\theta _{j} \}_{u_j \in M'_B},\tilde G_M)\right|,
\nonumber
\\
&\equiv& 2^{-2|M_A|-|\partial M_{AB}|-|M_B|} \left| \mathcal{Z} ( \{   s_i\}^{*}, \{ \theta _{j} \}^{*},\tilde G_M)\right|
\label{overlap2}
\end{eqnarray}
where we defined $\{ s_i \}^{*} \equiv \{ s_{i}\} _{M} \cup \{ 0 \}_{ v_i \in \partial M_{AB}} \cup \{s'_i \}_{M'} $ and $\{ \theta _j \}^{*}\equiv
\{\theta _{j} \}_{u_j \in M_B} \cup \{ -\theta _{j} \}_{u_j \in M'_B}$. 
We should note that $s_i$ and $s'_i$ 
take the same value but 
$\theta _j$'s sign is flipped on its copy $u'_j \in M'_B$.
From Eqs. (\ref{overlap1}) and (\ref{overlap2}),
\begin{eqnarray*}
P_{IQP}(\{ s_{i}\}_{i \in M}| \{ \theta _{j} \} , \{S_j \}, M )
&=&2^{-2|M_A|-|\partial M_{AB}|}
\left| \mathcal{Z} ( \{   s_i\}^{*}, \{ \theta _{j} \}^{*},\tilde G_M)\right|
\end{eqnarray*}
That is, the marginal distribution with respect to the set $M$ of the measured qubits is 
given by the normalized squared norm of the partition 
function of the Ising model defined by the merged graph $\tilde G_M$.
\hfill $\square$

The above theorem also indicates that 
the marginal distribution is equivalent 
to the square root of the joint probability 
of the {\sf IQP} circuit associated with the merged graph $\tilde G_M$,
weights $\{ \theta _j \}^{*}$ and the measurement outcomes $\{ s_i\}^{*}$:
\begin{eqnarray*}
P_{IQP}(\{ s_{i}\}_{i \in M}| \{ \theta _{j} \} , \{S_j \}, M )
=\left[ P_{IQP}(\{ s_i\}^*,\{\theta _j\}^*, \{ \mathcal{N}_{u_j} | u_j \in \tilde G_M\} )\right]^{1/2}.
\end{eqnarray*}

This indicates that
if the joint probability distributions of the {\sf IQP} circuits 
associated with a class of graphs
can be calculated efficiently,
and the class of graphs is closed 
under merging mentioned above,
then the marginal distributions of such a class of {\sf IQP} circuits
can also be calculated efficiently.
An example of such a class is planar graphs,
where the merged graph $\tilde G_{M^{(k)}}$ is also a planar graph
with an appropriately chosen measurement order such that $M^{(k)}$ is  
always connected.

Conditioned on the measurement outcome $\{ s_i \}_{i \in M}$ on 
the set $M$,
the probability of obtaining 
the next measurement outcome $s_k$ 
is calculated by using the Bayes rule as
\begin{eqnarray*}
p(s_k | \{ s_i \} _{i \in M})=\frac{
P_{IQP}(\{ s_ i\}_{i \in M\cup k}| \{ \theta _j \} , \{S_j\},  M \backslash k )
}{
P_{IQP}(\{ s_ i\}_{i \in M}| \{ \theta _j \} , \{S_j\},  M ) }.
\label{recursive}
\end{eqnarray*}
By denoting the set of all measured qubits after the $k$th measurements as $M^{(k)}$
(since there is no order in the measurements in {\sf IQP},
we can choose an arbitrary order of measurements for our convenience),
we can reconstruct the joint probability distribution of {\sf IQP} 
as follows:
\begin{eqnarray*}
P_{IQP}( \{ s_i \} | \{\theta _j\} , \{ S_j \})
= \prod _{k=1}^{n} p( s_{i_k} | \{s_i \}_{i \in M^{(k)}} ),
\end{eqnarray*}
where the $i_k$th qubit is measured at step $k$,
i.e., $\{i_k \} \cup M^{(k-1)} = M^{(k)}$.
If the marginal distribution, that is,
the Ising partition functions defined on $\tilde G_{M^{(k)}}$ can be calculated efficiently for all $M^{(k)}$
for a measurement order,
{\sf IQP} is classically simulatable at least in the weak sense.

Note that even if we can calculate the marginal distributions 
for an appropriately chosen measurement order, it is not sufficient to show strong simulatability
in a strict sense. 
In order to shown strong simulatability,
we have to show that arbitrary marginal distributions 
can be calculated efficiently.
In the next section,
we will see a classically simulatable class
based on planarity of the associated Ising models.
However, if we choose a wrong measurement order,
the merged graph results in a non-planar graph.
In such a case, the marginal distribution
is mapped into a partition function of an Ising model on a non-planar lattice,
which is hard to calculate~\cite{Barahona,Istrail,RaussendorfNonplanar}.
To clarify this situation, 
we say {\it almost strongly simulatable} if 
there exists a measurement order, and all marginal distributions 
with respect it can be calculated efficiently. 

\section{Classical simulatable classes of {\sf IQP}}
\label{sec4}
In general, exact calculation of partition functions 
of Ising models in the presence of magnetic fields 
is highly intractable in classical computer even on 2D planar lattice~\cite{Barahona,Jerrum93}.
The Ising models, to which we have mapped {\sf IQP}  in Sec.~\ref{sec3},
include the random $i\pi/2$ magnetic fields
depending on the output $\{ s_i\}$.
Thus one might think that 
we cannot find a nontrivial class
of {\sf IQP} that is classically simulatable.
This is, however, not the case.
Below we will show that
if the geometries of the graphs 
have some properties,
we can safely remove the magnetic fields 
renormalizing it into the coupling constants $\{ \theta _j\}$.

In this section, we will provide two classes of 
{\sf IQP} that are classically simulatable efficiently.
One is based on the sparsity of the commuting gates.
The other is based on the exact solvability of 
Ising models on the 2D planar lattices 
without magnetic fields~\cite{Kasteleyn,Barahona,Fisher}.
The former is strongly simulatable and the latter is 
at lest weakly simulatable and almost strongly simulatable.
In both cases, classical simulatability can be shown under arbitrary rotational angles $\{\theta _j\}$.

\subsection{Classical simulatability: sparse commuting circuits}
\begin{figure}[t]
\begin{center}
\includegraphics[width=0.9\textwidth]{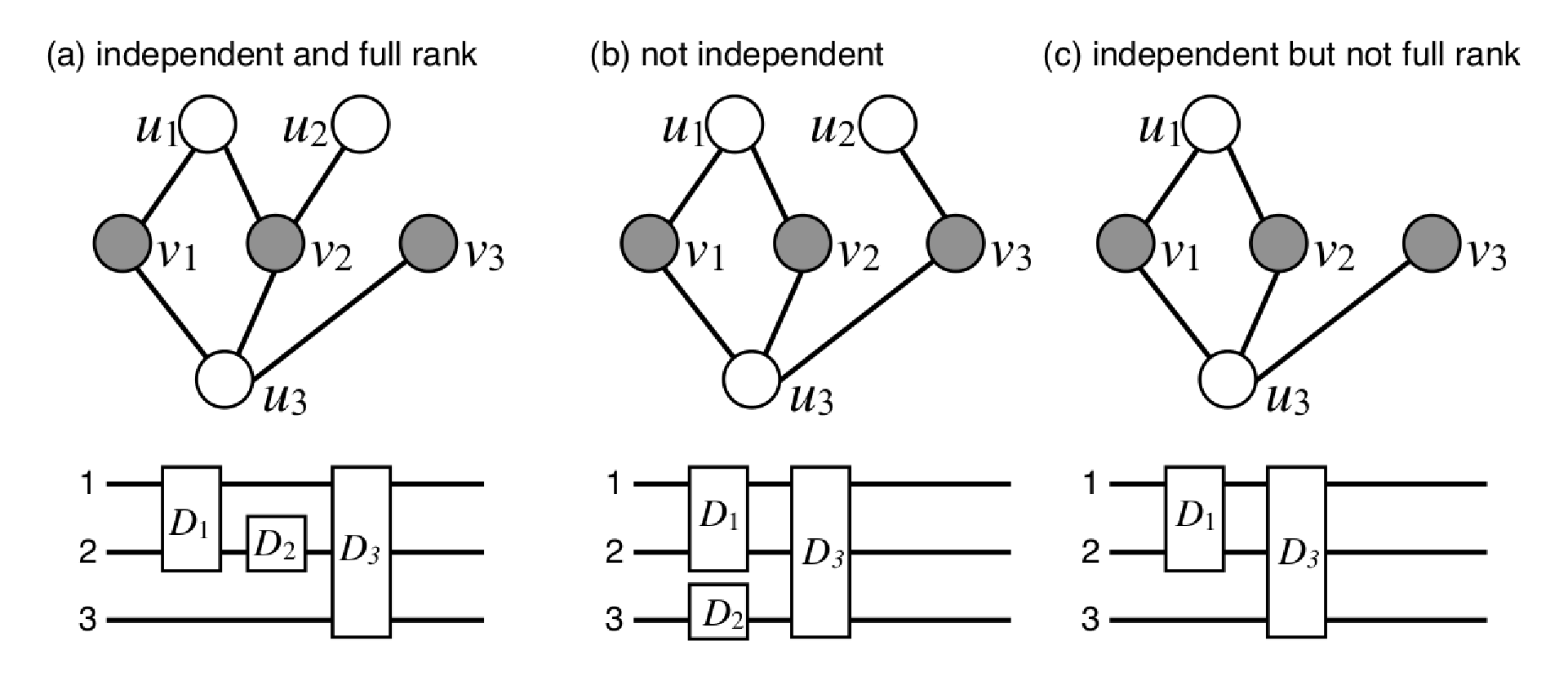}
\end{center}
\caption{Bipartite graph states (top) and associated commuting circuits (bottom). The white and gray shaded circles indicate qubits in $U_B$ and $V_A$, respectively. 
 (a) An independent and full rank bipartite graph. (b) A non-independent bipartite graph. (c) An independent but non-full rank bipartite graph.
 } 
\label{fig1}
\end{figure}

Let us define a $|V_A| \times |U_B|$ matrix $R$,
associated with the bipartite graph $G=(V_A \cup U_B,E)$,
such that $R_{v_i}^{u_j} = 1$ iff a vertex $v_i \in V_A$ 
is in $\mathcal{N}_{u_j}$, otherwise $R_{v_i}^{u_j}=0$.
We consider a class of bipartite graphs,
for which the row vectors of $R$ 
are linearly independent and full rank (and therefore $|V_A|=|U_B|$) 
in $\mathbf{Z}_2^{|U_B|}$
(later we will weaken the latter condition).
This condition implies that the column vectors of $R$ are also 
linearly independent and full rank.
We call such a bipartite graph as independent and full rank bipartite (IFRB) graph.
An example of an IFRB graph is depicted in Fig.~\ref{fig1} (a).

Now we consider the Ising model associated with 
an IFRB graph.
If we consider only computational basis, we can replace
the classical spin variable $\sigma$ with the Pauli $Z$ operator.
Therefore, we can rewrite the Ising Hamiltonian Eq. (\ref{Isingmodel})  
as 
\begin{eqnarray*}
\hat H(\{s_{i}\}, \{ \theta _{j} \},G) \equiv -  
\sum_{i}  \frac{i\pi}{2} s_{i} (1-Z _{v_i})
- \sum _{j} i  \theta _{j}\left(  
\bigotimes _{ v_i \in N_{u_j} } Z _{v_i} \right).
\end{eqnarray*}
Then the partition function is given by 
\begin{eqnarray*}
\mathcal{Z}(\{s_{i}\}, \{ \theta _{j} \},G) = {\rm Tr} \left[ e^{ - 
\hat H (\{s_{i}\}, \{ \theta _{j} \},G) } \right].
\end{eqnarray*}

Our main goal here is to
calculate $|\mathcal{Z}(\{s_{i}\}, \{ \theta _{j} \},G)| ^2$ exactly.
To this end, let us first consider the case $s_{i}=0$ for all $v_i$. 
In this case, there is no magnetic field, and hence
we can transform the Hamiltonian into an interaction-free Ising model
by virtue of the properties of the IFRB graph.
\begin{lemm}[Mapping to interaction-free Ising model]
\label{renoma1}
For any Ising model associated with an IFRB graph,
there exists a unitary operator $W$ that
transforms $\hat H(\{0\}, \{ \theta _{j} \},G)$ to interaction-free
Ising Hamiltonian:
\begin{eqnarray*}
W\hat H (\{0\}, \{ \theta _{j} \},G)W^{\dag}= \sum _{j} i \theta _j Z_{v_j}
\end{eqnarray*}
\end{lemm}
{\it Proof}:
Since the column vectors of $R$ 
are independent and full rank,
we can transform the matrix $R$ to the identity matrix 
by using the Gauss-Jordan elimination method.
Since the matrix $R$ defines the graph
and the Hamiltonian,
the Gauss-Jordan elimination can be viewed as
a transformation of the graph and the corresponding Hamiltonian.
The graph associated with the identity matrix 
consists of pairs of vertices $(v_i,u_i)$
connected by edges. Since each vertex in $U_B$
is always connected only one vertex in $V_A$,
the corresponding Ising Hamiltonian is interaction-free.

Each process in the Gauss-Jordan elimination for the matrix $R$ 
can be implemented on the Hamiltonian 
by conjugations of controlled-Not (CNOT) and 
swapping gate operations.
The CNOT gate from the $i$th to the $j$th qubits is equivalent to 
adding the $j$th row vector to the $i$th one on the matrix $R$.
The swapping gate exchanges the labels $\{ v_i\}$ of the vertices. 
Thus there exists a unitary operator $W$
consisting of swapping and CNOT gates 
such that $W \hat H(\{ 0\} , \{ \theta _j \} , G) W^{\dag}
= \sum _{j} i \theta _j Z_{v_j}$.
\hfill $\square$

For example, in the case of the IFRB graph shown in Fig.~\ref{fig1} (a),
the set of operators in the Hamiltonian is given by 
$\{ Z_{v_1}Z_{v_2}, Z_{v_1} Z_{v_2} Z_{v_3}, Z_{v_2} \}$.
This can be mapped to $\{ Z_{v_1}, Z_{v_2} ,Z_{v_3}\}$
by using the unitary operator $W=S^{wap}_{v_2,v_3} \Lambda (X) _{v_1,v_3} \Lambda (X)_{v_2,v_1}$,
where $S^{wap}_{v_i,v_j}$ is the swapping operation between qubits $v_i$ and $v_j$.

By using such a $W$, 
the partition function can be calculated as
\begin{eqnarray*}
\mathcal{Z}(\{s_{i}\}, \{ \theta _{j} \},G) &=& {\rm Tr} \left[ e^{ - \hat{H} (\{s_{i}\}, \{ \theta _{j} \}) } \right]
\\
&=&
{\rm Tr} \left[W  e^{ - W^{\dag}\hat H (\{s_{i}\}, \{ \theta _{j} \})W} W^{\dag} \right]
\\
&=& 2^{ |U_B|} \prod _{u_j} \cos  \theta _{j} .
\end{eqnarray*}
Thus the probability of obtaining $\{ s_{i} = 0 \}$ is computed as
\begin{eqnarray*}
P_{\rm IQP}(\{ s_{i}=0 \} | \{ \theta _{j} \},\{S_j\}) = \Big( \prod _{u_j} \cos  
\theta _{j}\Big)^2. 
\end{eqnarray*}
Since the joint probability is factorized for each $\theta_j$,
we can easily calculate its marginal distribution (without using Theorem~\ref{Main2} in this case).

Next we extend the above result to the general measurement outcomes $\{ s_{i} \}$.
This is done by renormalizing the random $i\pi /2$ magnetic fields
into the coupling constants as follows.
\begin{lemm}[Renormalization of $i \pi/2$ magnetic fields]
\label{renoma2}
For any {\sf IQP} associated with an IFRB graph,
we can find a bit string $\{c_{u_j}\}$
such that
\begin{eqnarray*}
P_{\rm IQP}(\{ s_{i} \} | \{  \theta _{j}  \})
=P_{\rm IQP}(\{ s_{i}=0 \} | \{ \tilde  \theta _{j} \}),
\end{eqnarray*}
with $\tilde \theta _{j} \equiv  \theta _{j} + c_{u_j} \pi /2$.
\end{lemm}
{\it Proof}:
Let us consider the corresponding {\sf MBIQP}.
From the definition of {\sf MBIQP},
\begin{eqnarray*}
P_{MBIQP} (\{ m_{v_i} \},\{ m_{u_j}\} | \{ \theta _j \}, G)
&=& \left| \bigotimes _{v_i \in V_A} \langle +_{m_{v_i}} |
\bigotimes _{u_j \in U_B} \langle \theta _{j,m_{u_j}} |    | G\rangle \right|^2
\\
&=&
\left|  \langle +_{0} |^{\otimes |V_A|} F(\{ m_{v_i}\})
\bigotimes _{u_j \in U_B} \langle \theta _{j,m_{u_j}} |  | G\rangle \right|^2,
\end{eqnarray*}
where $F(\{m_{v_i} \})\equiv\bigotimes _{v_i \in V_A} Z_{v_i} ^{ m_{v_i} }$. 
Since the row vectors of $R$ are independent and full rank,
we can find a vector $c_{u_j}$ in ${\bf Z}_2^{|U_B|}$ such that 
$m_{v_i} = \sum_{u_j}R_{v_i}^{u_j} c_{u_j}$ for any $\{ m_{v_i} \}$.
By using this vector $c_{u_j}$,
we obtain the following equality,
\begin{eqnarray*}
\prod _{u_j \in U_B } (X_{u_j}K_{u_j})^{c_{u_j}}=
\prod _{u_j \in U_B } \left ( \prod _{v_i\in \mathcal{N}_{u_j}}Z_{v_i} \right) ^{c_{u_j}} = F(\{ m_{v_i} \}).
\end{eqnarray*}
By using this and the fact that $K_{u_j}$ stabilizes
$|G\rangle$, we obtain 
\begin{eqnarray*}
P_{MBIQP} (\{ m_{v_i} \},\{ m_{u_j}\} | \{ \theta _j \}, G)
&=& 
\left|  \langle +_{0} |^{\otimes |V_A|} 
\bigotimes _{u_j \in U_B} \langle \theta _{j,m_{u_j}} | \left( \prod _{u_j \in U_B } X_{u_j}^{m_{v_i}} \right)| G\rangle \right|^2
\\
&=&
\left|  \langle +_{0} |^{\otimes |V_A|} 
\bigotimes _{u_j \in U_B} \langle \tilde \theta _{j,m_{u_j}} || G\rangle \right|^2
\\
&=&P_{MBIQP} (\{ \tilde s_{v_j}=0 \}, \{ m_{u_j}\}| \{ \tilde \theta _{j}\},G ),
\end{eqnarray*}
where $\tilde \theta _{j} \equiv  \theta _{j} + c_{u_j} \pi /2 $.
Specifically, if we consider the case $m_{u_j}=0$,
we obtain that 
\begin{eqnarray*}
 P_{IQP}(\{s_i\}| \{ \theta _j \} , \{ S_j\})
&=&2 ^{ |U_B|}P_{MBIQP}(\{s_{v_i}\}, \{ m_{u_j}=0 \}| \{ \theta _j \} , G)
\\
&=&2 ^{ |U_B|}P_{MBIQP}(\{s_{v_i}=0\}, \{ m_{u_j}=0 \}| \{ \tilde \theta _j \} , G)
\\
&=&P_{IQP}(\{s_i=0\}| \{ \tilde \theta _j \} , \{ S_j\}).
\end{eqnarray*}
\hfill $\square$

Let us consider the example shown in Fig.~\ref{fig1} (a) again.
For instance, if $\{s_{v_i}\}=\{0,0,1\}$,
$F(\{0,0,1\})= Z_{v_3}$, and $\{c_{u_1}=1, c_{u_2}=0,c_{u_3}=1\}$. By multiplying 
the stabilizer operators of the graph state with respect to the $4$th and $6$th vertices,
we obtain another stabilizer operator $(X_{u_1} Z_{v_1} Z_{v_2})(X_{u_3} Z_{v_1} Z_{v_2} Z_{v_3})= X_{u_1} X_{u_3} Z_{v_3}$.
Thus the action of $F(\{ 0,0,1\})$ is equivalent to that of $X_4 X_6$,
which rotates the angles $\theta _{u_1}$ and $\theta _{u_3}$ by $\pi /2$.

By combining Lemma~\ref{renoma1} and Lemma~\ref{renoma2},
we can show classical simulatability of {\sf IQP}
associated with IFRB graphs.
\begin{theo}[Classical simulatability: sparse circuits]
\label{sparseIQP}
{\sf IQP} associated with an IFRB graph is classically simulatable.
\end{theo}
{\it Proof}:
From Lemma~\ref{renoma1} and~\ref{renoma2},
we can calculate $P_{\rm IQP}(\{ s_{i} \} | \{  \theta _{j}  \})$
exactly for an IFRB graph
including its arbitrary marginal distributions. 
Thus such a class of {\sf IQP}
is classically simulatable for arbitrary angles $\{ \theta _{j}\}$ in the strong sense.
\hfill $\square$

Finally, we slightly weaken the condition, full rank.
Even if the column vectors of $R$ is not full rank, i.e., $|U_B|< |V_A|$ [as shown in Fig.~\ref{fig1} (c)], 
there exist $W$ such that transforms the many-body Ising Hamiltonian 
to interaction-free Ising Hamiltonian as long as the column vectors of $R$ are independent.
Such a class of graphs are called independent bipartite (IB) graphs.
In this case, the existence of $c_{u_j}$ for all $\{m_{u_j} \}$
is not guaranteed, and hence we have to find another way 
to deal with this situation.

To settle this, we add ancilla vertices $u_{j'} \in U_{B'}$
to the set $U_B$ in such a way that $R_{v_i}^{u_j}$ ($u_j \in U_B \cup U_{B'}$)
has full rank [The 5th qubit in Fig.~\ref{fig1} (a) can be viewed as the ancilla qubit
for the non-full rank graph in Fig~\ref{fig1} (c)]. 
Due to Theorem~\ref{sparseIQP},  
we can exactly calculate the probability for the slightly enlarged problem,
$P_{\rm IQP}(\{ s_{i} \} | \{  \theta _{j}  \} \cup \{ \theta _{j'} \})$.
Then, the probability $P_{\rm IQP}(\{ s_{i} \} | \{  \theta _{j}  \} )$, with which we want to sample 
$\{ s_{i} \}$,
can be obtained by considering a specific case $\theta _{j'}=0$ for all $u_{j'} \in U_{B'}$, i.e.,
\begin{eqnarray*}
P_{\rm IQP}(\{ s_{i} \} | \{  \theta _{j}  \} \cup \{ \theta _{j'}=0 \}  ) = 
P_{\rm IQP}(\{ s_{i} \} | \{  \theta _{j}  \} ).
\end{eqnarray*}
A representative example of classically simulatable {\sf IQP} circuits 
are depicted in Fig.~\ref{fig1} (a) and (c).
If we restrict ourselves into two-body Ising models (i.e., $|S_j|=2$), the meaning of independence
becomes clear; independence means that the lattice does not
contain any loop, such as Ising models on one-dimensional chain or tree graphs. 
Thus {\sf IQP} with two-qubit commuting gates
whose interaction geometry does not contain any loop can be efficiently simulated in the strong sense.
In order to avoid the present class of classically simulatable {\sf IQP},
the {\sf IQP} circuits that consist of at least $n$ ($=|V_A|$) commuting gates 
acting on different subsets $\{S_j\}$ of qubits 
are sufficient.

\subsection{Classical simulatability: planar-{\sf IQP}}
Classical simulatability in the previous case
is based on the sparsity of the commuting gates,
where at most only $n-1$ commuting gates are included.
In such a case we can calculate the partition functions
without using Theorem~\ref{Main2}.
Next we will provide another classically simulatable 
class of {\sf IQP},
that includes commuting gates much more than $n$.
Specifically, we will show below that
{\sf IQP} with two-qubit commuting gates acting on nearest-neighbor two qubits on the 2D planar graphs,
which we call planar-{\sf IQP}, is classically simulatable
almost in the strong sense.
That is, the probability distribution of the output
and its marginal distribution for an appropriately chosen measurement order
can be calculated efficiently.
To this end,
we first show, by using properties of the graph states, that 
for two-body Ising models we can always remove 
the random $i \pi /2$ magnetic fields by 
appropriately renormalizing their effects into
coupling constants $\{ \theta_j \}$.
This allows us to map planar-{\sf IQP}
to two-body Ising models without magnetic fields.
Then we utilize Theorem~\ref{Main2} and exact solvability of 
two-body Ising models on planar lattices to construct 
an efficient classical simulation of {\sf IQP}.

Consider a planar bipartite graph $G$ 
with $|S_j|=2$, that is, every vertex $u_j \in U_B$
are connected with just two vertices $v_i \in V_A$.
The weights $\{ \theta _j\}$ are arbitrary.
For simplicity, we assume 
that $G$ is connected.
Let us consider properties of 
the graph state associated with such a planar bipartite graph $G$.

\begin{remk}[Property of Graph states 1]
\label{property1}
For any connected bipartite graph $G$ with $|S_j|=2$ for all $j$,
the associated graph state $|G\rangle$ is subject to
the following property:
\begin{eqnarray*}
\left( \prod _{v_i \in V_A} \langle + _{m_{v_i}} |  \right)
|G\rangle = 0
\end{eqnarray*}
for any $\{m_{v_i}\}$ such that $\bigoplus _{v_i \in V_A} m_{v_i} =1$.
Here the addition is taken modulo two.
\end{remk}
{\it Proof}:
The bipartite graph state is 
stabilized by 
\begin{eqnarray*}
\prod_{v_i \in V_A} \left( X_{v_i} \prod_{u_j \in \mathcal{N}_{v_i}} Z_{u_j} \right) 
= \prod_{v_i \in V_A} X_{v_i},
\end{eqnarray*}
and hence $\left(\prod_{v_i \in V_A} X_{v_i} \right)|G\rangle =|G\rangle$.
By using this, we obtain
\begin{eqnarray*}
\left( \prod _{v_i \in V_A} \langle + _{m_{v_i}} |  \right)
|G\rangle = \left( \prod _{v_i \in V_A} \langle + _{m_{v_i}} |  \right)
\left(\prod_{v_i \in V_A} X_{v_i} \right)
|G\rangle 
=\left( \prod _{v_i \in V_A} \langle + _{m_{v_i}} |  \right)
(-1)^{\bigoplus _{v_i \in V_A} m_{v_i}}
|G\rangle .
\end{eqnarray*}
Thus if $\bigoplus _{v_i \in V_A} m_{v_i} =1$, then $\left( \prod _{v_i \in V_A} \langle + _{m_{v_i}} |  \right)
|G\rangle =0$.
\hfill $\square$

Thus we only consider the case $\bigoplus _{v_i \in V_A} m_{v_i}=0$,
that is, the number of vertices with $m_{v_i} =1$ is even. 
In such a case, we can show that 
modifying the coupling constants $\{\theta _j\}$ appropriately
as follows can renormalize
$i \pi/2$ magnetic fields.

\begin{figure}[htbp]
\begin{center}
\includegraphics[width=0.9\textwidth]{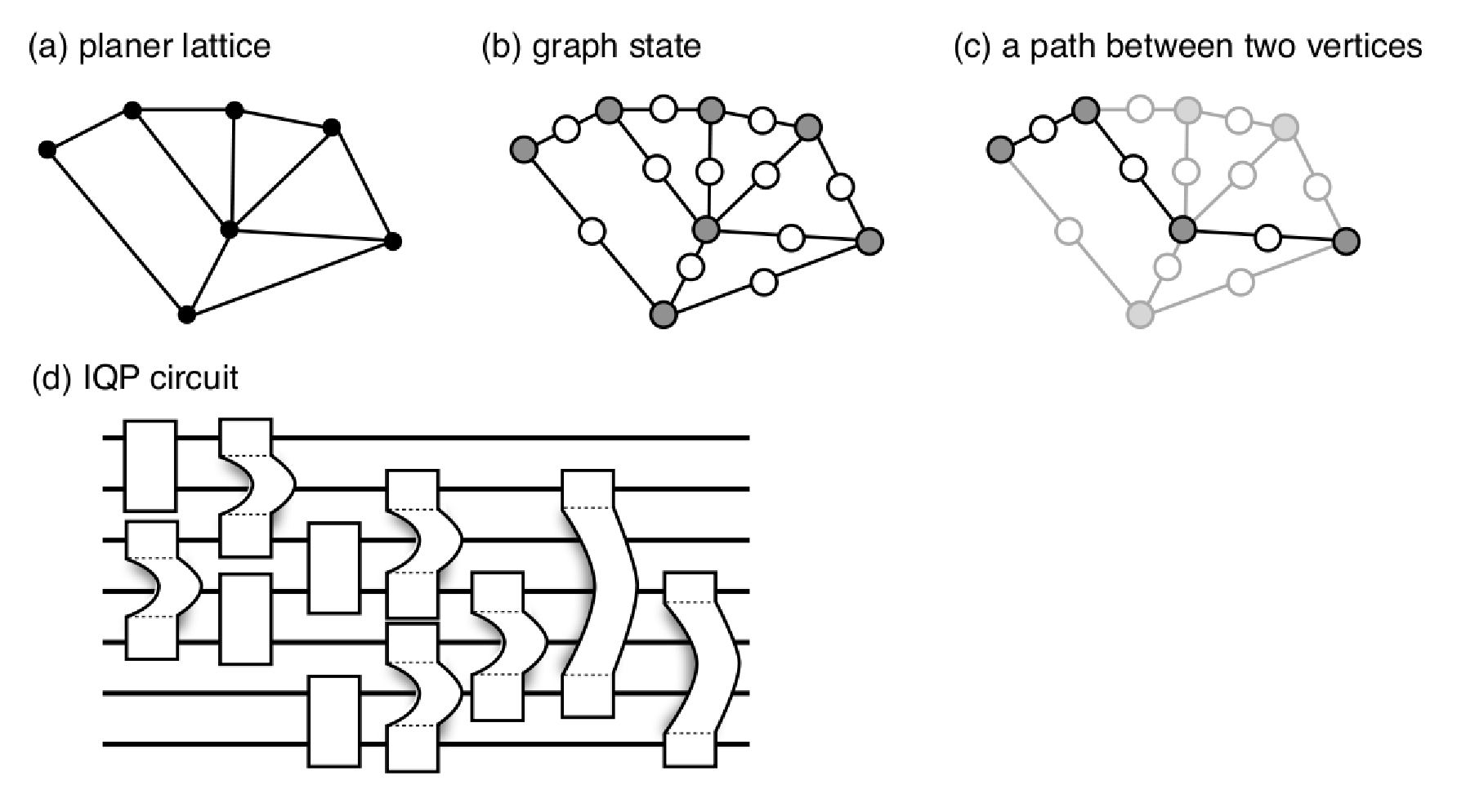}
\end{center}
\caption{(a) A planar lattice. (b) An associated bipartite graph state,
where gray and white circles denote qubits in $V_A$ and $U_B$, respectively.
(c) A path between a pair of qubits in $V_A$. (d) The corresponding commuting circuit.
 } 
\label{fig2}
\end{figure}

\begin{remk}[Property of Graph states 2]
\label{property2}
For any {\sf IQP} associated with 
a connected bipartite graph $G$ with $|S_j|=2$ for all $j$,
by appropriately choosing $\{ \tilde \theta _j\}$,
\begin{eqnarray*}
P_{IQP}(\{s_i \} | \{ \theta _j\}, \{ S_j\} )
=P_{IQP}(\{ s_i=0 \} | \{ \tilde \theta _j\}, \{ S_j\} ),
\end{eqnarray*}
where $\{ s_i=0 \}$ means that $s_i=0$ for all $i$.
Equivalently, 
for the corresponding Ising models, we have
\begin{eqnarray*}
H(\{s_i\}, \{ \theta _j\}, G) =H(\{s_i=0\}, \{ \tilde \theta _j\}, G) ,
\end{eqnarray*}
that is, the random $i \pi /2$ magnetic fields 
can be renormalized into the coupling constants $\{\tilde \theta _j \}$.
\end{remk}
{\it Proof}:
Consider the graph state $|G\rangle$.
Due to Remark~\ref{property1}, 
the number of $\tilde s_i=1$ is always even.
The graph is connected.
Thus we can always make pairs of vertices $v_i \in V_A$
of $m_{v_i} =1$. Apparently this can be done in polynomial-time,
since arbitrary paring is allowed.
Let us denote such a pair as $(v_k \sim v_{k'})$
and a set of vertices on a path (arbitrarily) connecting them as ${\rm path}(v_k \sim v_{k'})$.
The graph state is stabilized by  
\begin{eqnarray*}
\prod _{u_j \in {\rm path}(v_k \sim v_{k'}) \cap U_B}
K_{u_j} =
Z_{v_k} \left( \prod_{u_j \in {\rm path}(v_k \sim v_{k'}) \cup U_B} X_{u_j} \right) Z_{v_{k'}},
\end{eqnarray*}
[see Fig.~\ref{fig2} (b) and (c)].
By using this fact,
we can obtain
\begin{eqnarray*}
&&\left(\bigotimes _{v_i \in V_A} \langle +_{m_{v_i} } |\right)
\left( \bigotimes _{u_j \in U_B} \langle \theta _{j,m_{u_j}} | \right)
 | G\rangle
\\
&=&\left(\bigotimes _{v_i \in V_A} \langle +_{m_{v_i} } |\right)
\left( \bigotimes _{u_j \in U_B} \langle \theta _{j,m_{u_j}} | \right)
\left[ Z_{v_k} \left( \prod_{u_j \in {\rm path}(v_k \sim v_{k'}) \cup U_B} X_{u_j} \right) Z_{v_{k'}} \right] |G\rangle
\\
&=&
\left(\bigotimes _{v_i \in V_A} \langle +_{m_{v_i} \oplus \delta _{v_i,v_k}
\oplus \delta _{v_i,v_{k'}} } |\right)
\left( \bigotimes _{u_j \in U_B} \left\langle  \theta  _{j,m_{u_j}\bigoplus _{u_{j'} \in {\rm path}(v_k \sim v_{k'})} \delta _{u_j,u_{j'}}} \right| \right) |G\rangle.
\end{eqnarray*}
By doing this repeatedly for 
all pairs of $m_{v_i}=1$, i.e.,
a perfect matching of $m_{v_i}=1$ vertices,
we can transform all $m_{v_i}=1$ to $m_{v_i}=0$.
Let us define an arbitrary perfect matching $\mathcal{M}$ of vertices of $m_{v_i}=1$ 
and a set ${\rm path}(\mathcal{M})$ of paths of the matching $\mathcal{M}$.
By denoting the addition modulo two
over $u_{j'}$s on all these paths by $\bigoplus _{ u_{j'} \in{\rm path} (\mathcal{M})}$,
the renormalized coupling constant
is given by
\begin{eqnarray*}
\tilde \theta _j  = \theta _j + \left( \bigoplus _{u_{j'} \in{\rm path}(\mathcal{M})} \delta _{ u_j, u_{j'}}  \right) \pi /2.
\end{eqnarray*}
Then we obtain
\begin{eqnarray*}
\left(\bigotimes _{v_i \in V_A} \langle +_{m_{v_i} } |\right)
\left( \bigotimes _{u_j \in U_B} \langle \theta _{j,m_{u_j}} | \right)
 | G\rangle
 =
\langle +_{0 } |^{\otimes |V_A|}
\left( \bigotimes _{u_j \in U_B} \langle \tilde \theta _{j,m_{u_j}} | \right)
 | G\rangle.
\end{eqnarray*}
This leads that
\begin{eqnarray*}
P_{IQP} (\{s_i\} | \{\theta _j\}, \{ S_j\})&=&
2^{|U_B|}P_{MBIQP}( \{ m_{v_i}\},\{ m_{u_j}=0 \}| \{ \theta _j \} , G)
\\
&=&2^{|U_B|}P_{MBIQP}( \{\tilde s _{v_i}=0\},\{ m_{u_j}=0 \}| \{ \tilde \theta _j \} , G)
\\
&=&P_{IQP} (\{s_i =0\} | \{\tilde \theta _j\}, \{ S_j\})
\end{eqnarray*}
\hfill $\square$

Note that in the proofs of the properties of graph states
with $|S_j|=2$, we did not use the planeness of the graph.
Thus Remark~\ref{property1} and Remark~\ref{property2} hold
even for nonplanar graphs as long as $|S_j |=2$ for all $j$.
Accordingly, we can always remove the 
random $i \pi /2$ magnetic fields of arbitrary two-body Ising models
by appropriately renormalizing them into the two-body coupling 
constants.

Interestingly, these properties of the graph states
are closely related to the properties of anyonic excitations
on surface codes with a smooth boundary~\cite{Kitaev}.
On the graph state with $|S_j|=2$ for all $j$, 
if one project the qubits in $V_A$ by $|+\rangle^{\otimes |V_A|}$,
we obtain the surface code state defined on a lattice $\mathcal{L}$,
where vertex and edge corresponds to vertices in $V_A$ and $U_B$ of $G$
respectively,
and a qubit is assigned on each edge.
This can be confirmed as follows. 
The post-measurement state is stabilized by 
$\prod _{u_j \in \mathcal{N}_{v_i}} Z_{u_j} \equiv A_{v_i}$ for all $v_i$.
Furthermore, 
for all faces $f$ of the lattice $\mathcal{L}$,
$  \prod _{u_j \in \partial f} K_{u_j} =\prod _{u_j \in  \partial f} X_{u_j}\equiv B_f$
stabilizes the post-measurement state,
where $\partial f$ is the set of the edges that are boundary of the face $f$.
These two types operators are called 
star and plaquette operators in Ref.~\cite{Kitaev}. 
The post-measurement state 
or equivalently the surface code state 
is the ground state of the Hamiltonian,
so-called Kitaev's toric code Hamiltonian,
\begin{eqnarray*}
H= -J \sum _{i} A_i  - J \sum _{f} B_f.
\end{eqnarray*}
A projection by $|-\rangle _{v_i} $ results in
the eigenvalue $-1$ of the star operator at vertex $v_i$,
which corresponds to the anyonic excitation in the Kitaev model.
Then Remark~\ref{property1} indicates that the parity of anyonic excitations
is always even. They are created and annihilated in pairs. Remark~\ref{property2} corresponds a way to annihilate the pairs of the anyonic excitations. The trajectory of anyonic excitations in the annihilation 
process corresponds to ${\rm path}(\mathcal{M})$.

Now we are ready to show that classical simulatability 
of {\sf IQP} consisting of 2D nearest-neighbor two-qubit commuting gates. 
\begin{theo}[Classical simulatability: planar-{\sf IQP}]
\label{freefermion}
planer-{\sf IQP} consisting of two-qubit commuting gates 
acting on nearest-neighbor qubits on the 2D planar graphs
is classically simulatable almost in the strong sense.
\end{theo}
{\it Proof}:
According to Theorem~\ref{Main1},
the joint probability distribution of planar-{\sf IQP}
can be calculated from a two-body Ising partition function on a planar lattice.
Since the graph $G$ is a planar bipartite graph,
we can easily find an order of measurements such that
$\tilde G_{M^{(k)}}$ is also planar at any measurement step $k$.
(Any order of measurements such that the subgraph $G_{M^{(k)}}$ becomes a connected graph
for all $k$ can be utilized.) 
Due to Theorem~\ref{Main2}, 
the marginal distributions are 
also given as Ising partition functions on planar lattices.
Furthermore, in the merged graph, the vertices $u_j \in M_B^{(k)} \cup {M'}^{(k)}_{B}$
are connected with just two vertices, i.e., $|\mathcal{N}_{u_j}|=2$.
For such Ising models, by using Remark~\ref{property1} and Remark~\ref{property2},
the random magnetic $i \pi /2$ fields can be renormalized into the coupling constants 
$\{ \theta \} \rightarrow \{ \tilde \theta _j \}$.
Thus all marginal distributions
can be calculated from the two-body Ising partition functions 
on planar lattices without magnetic fields.
On the other hand,
it is well known that
the partition function of two-body Ising models
on planar lattices without magnetic fields can
be calculated efficiently by expressing 
them as the Pfaffians~\cite{Kasteleyn,Fisher,Barahona}. 

Thus we conclude that 
{\sf IQP} of this class can be simulated efficiently
almost in the strong sense,
which is sufficient for an efficient weak simulation with a recursive method.
\hfill $\square$

Note that a similar argument is also made in Ref.~\cite{BravyiRaussendorf} 
by considering classical simulatability of MBQC on the planar surface codes~\cite{Kitaev}.
Indeed, as mentioned before, 
if we apply the projection by $|+ \rangle ^{ \otimes |V_A|}$
on the bipartite planar graph state with $|S_j|=2$,
we obtain an unnormalized planar surface code state consisting of the qubits on $U_B$.
The effect of $m_{v_i}=1$ (i.e., the projection by $|+_{1}\rangle$) 
can be renormalized into
the coupling constants $\{\theta _j \} \rightarrow \{ \tilde \theta _j\}$,
where an arbitrary perfect matching is chosen 
as shown in Remark~\ref{property2}.
Thus we may construct an alternative proof of Theorem~\ref{freefermion}
without using Theorem~\ref{Main2}.
However, Theorem~\ref{Main2},
employing the properties of the graph states, is much straightforward and simple
for our purpose.
Furthermore, Theorem~\ref{Main2} is valid not only for
the case with $|S_j|=2$, but also
the general cases,
which cannot be regarded as MBQC on the planar surface codes.

While we have shown planar-{\sf IQP} is almost strongly simulatable,
it seems not to be strongly simulatable in the strict sense.
Suppose that we choose a measurement order $\{ M^{(k)}\}$
such that any subgraph $G^{M^{(k)}}$ consists of multiple disjoint subgraphs.
In such a case, the merged graph becomes a non-planar graph 
of a higher genus. 
The Ising partition functions on lattices of a higher genus
are hard to calculate in general~\cite{Barahona,Istrail,RaussendorfNonplanar}.
There seems to be an intermediate class of classical simulation,
which we named {\it almost strongly simulatable}, between 
strongly simulatable (in the strict sense) and weakly simulatable.

The Pfaffian is the square root of the determinant,
and hence the probability distribution of planar-{\sf IQP}
is given by the determinant of an appropriately defined complex matrix.
The determinant appears in 
the probability distribution of fermions scattered
by fermionic linear optical unitary operators.
Thus the present classical simulatable class of {\sf IQP}
is regarded as a {\sf F{\footnotesize ERMION}S{\footnotesize AMPLING}}.

Important implications of Theorem~\ref{freefermion} are twofold.
One is that planar-{\sf IQP} can generate highly 
entangled state but its output is classically simulatable 
almost in the strong sense.
This is also the case for the Clifford circuits and match gates,
which generate genuinely entangled states but are classically simulatable
~\cite{NielsenChuang,matchgates0,matchgates1,matchgates2,matchgates3}.
Secondary,
if single-qubit rotations are added to planer-{\sf IQP},
it becomes universal-under-postselection,
whose weak simulation is intractable unless the {\sf PH} collapses to the third level.
Thus single-qubit rotations take a quite important rule for
{\sf IQP} to be classically intractable.
Indeed, single-qubit rotations make a drastic change
of computational complexity from almost strongly simulatable to 
not simulatable even in the weak sense.

We would like to note that a similar result
is also obtained
in a rather different situation~\cite{WeakSim}.
He showed that Toffoli-Diagonal circuits,
which include quantum Fourier transformation for 
Shor's factorization algorithm,
can be efficiently simulated if there is no 
basis change at the final round before the
the computational basis measurements.
Thus single-qubit rotations also play a very important role
for the Toffoli-Diagonal circuits to be classically intractable.

Another consequence of Theorem~\ref{freefermion}
lies in the context of experimental verification of quantum benefits.
When we utilize {\sf IQP} for the purpose of 
experimental verification of quantum benefits,
we have to avoid planar-{\sf IQP},
since a malicious quantum device can cheat
experimentalists
by classically sampling the results instead of implementing the {\sf IQP} circuit. 
At the same time, 
the existence of efficient classical simulation for planar-{\sf IQP}
implies that
checking the correctness of experiments of this class is much easier.
Thus when experimentalists realize {\sf IQP},
they should, at least, 
try to implement planar-{\sf IQP}, 
since its correctness can be easily checked.
It might be possible to efficiently ensure,
under a plausible assumption,
that two-qubit commuting gates are implemented appropriately, 
since experimental devices 
are usually well known and not so malicious.
Hopefully, 
classical intractability of quantum devices
may be verified
by an efficient experimental verification of planar-{\sf IQP}
combined with other efficient witness or plausible assumptions~\cite{Tamate}.

Moreover, planar commuting circuits can generate 
an interesting class of entangled states,
called weighted graph states~\cite{GraphState}.
The constructed classical simulation 
would be useful to check an experimental 
preparation of such states efficiently.

\section{Hardness of approximating Ising partition functions}
\label{sec5}
In this section, we 
utilize the established relationship between
{\sf IQP} and Ising partition functions
in an opposite direction;
by considering universal-under-postselection instances of 
{\sf IQP},
we show that
a multiplicative approximation of Ising partition functions 
with almost all imaginary coupling constants is \#{\sf P}-hard
even on planar lattices with a bounded degree.
Note that this argument based on universality-under-postselection
and post-{\sf BQP} = {\sf PP} have been already utilized to show 
\#{\sf P}-hardness of approximating the permanent~\cite{AaronsonLinear} and the Jones polynomial~\cite{Kuperberg}.

\begin{theo}[Hardness of approximating imaginary Ising partition functions]
\label{FPRASHard}
A multiplicative approximation of Ising partition functions 
with almost all imaginary coupling constants is \#{\sf P}-hard
even on planar lattices with a bounded degree.
Thus if there exists a fully polynomial-time classical approximation scheme,
the {\sf PH} collapses completely.
\end{theo}

{\it Proof:}
We consider {\sf IQP} with a homogeneous rotational angle $\theta$.
As shown in Ref.~\cite{IQP},
{\sf IQP} associated with a bounded-degree planar graph with $|S_j| \leq 2$ 
is universal-under-postselection
when the homogeneous rotational angle is given by $\theta=\pi/8$.
Thus a multiplicative approximation of the Ising partition functions 
with the homogeneous coupling constant $i \theta = i \pi/8$ is \#{\sf P}-hard
due to Theorem~\ref{Main1} and Remark~\ref{StrongpostBQP}.
The same result holds not only $i\theta  =i \pi/8$
but also $i\theta = i(2l+1) \pi/(8m)$ for integers $l$ and $m$.

Suppose the homogeneous coupling
is given by an irrational angle i.e., $\theta = 2\nu  \pi$
with $\nu \in [0,1) $ being an irrational number.
Let $m$ be an integer.
Since $2 m\nu \pi $ ($\textrm{mod } 2\pi$) is distributed 
in a uniform fashion,
we can find an approximation of $\pi/8$ with an additive error $\epsilon$
with some integer $m = \mathcal{O}(1/\epsilon)$~\cite{NielsenChuang}.
Accordingly the commuting gates $D(2\nu \pi , S_j)^{m} =D(2m \nu \pi , S_j)$
is sufficiently close to the rotation $D(\pi/8,S_j)$
in the sense of an appropriately defined distance such as 
the diamond norm~\cite{Diamond}.
In the present case, the erroneous rotation $D(\pi/8 + \epsilon, S_j)$ is unitary, 
and hence the diamond norm is equivalent to 
the square of the operator norm,
which is given by
\begin{eqnarray*}
|| D (\pi/8 , S_j) [ I - D (\epsilon ,S_j) ] ||^2
=|| I - D (\epsilon ,S_j)  ||^2 = 2 (1-\cos \epsilon) = O(\epsilon).
\end{eqnarray*}
If a set of instances of {\sf IQP} is universal-under-postselection,
post-{\sf IQP} can simulate universal fault-tolerant quantum computation.
If the error $\epsilon$ is sufficiently smaller than
the threshold value of fault-tolerant quantum computation~\cite{DoritBenor,RaussendorfPhD,Nielsen},
we can reliably simulate universal quantum computation (i.e., {\sf BQP})
and moreover {\sf PP} with the help of postselection.
(See Ref.~\cite{Tamate} for an application of the fault-tolerance theory
to the postselection argument.)
Thus {\sf IQP} with almost all rotational angles 
is universal-under-postselection.
This fact and Remark~\ref{StrongpostBQP} lead that
a multiplicative approximation of the Ising partition functions is \#{\sf P}-hard
for almost all imaginary coupling constants even on planar lattices with a bounded degree.
\hfill $\square$

The above result indicates that 
almost all imaginary Ising partition functions 
are substantially hard to calculate even in the approximated case
with a multiplicative error.
This result contrasts with 
the existence of a FPRAS
in the ferromagnetic cases with magnetic fields shown by Jerrum and Sinclair~\cite{Jerrum93}
and antiferromagnetic cases on a sort of lattices 
shown by Sinclair, Srivastava, and Thurley ~\cite{Sinclair}.
In these cases, 
an exact calculation is \#{\sf P}-hard but its approximation with a multiplicative error is easy.
On the other hand, 
as noted in Remark~\ref{StrongpostBQP},
\#{\sf P}-hardness associated with post-{\sf BQP} = {\sf PP} theorem
is also holds in the approximated case automatically. 

With the random magnetic fields, approximation of ferromagnetic Ising 
partition functions below a certain critical temperature
belongs, under an approximation-preserving reduction,
to a class \#{\sf BIS}, which is defined as a counting problem of 
the number of independent sets of a bipartite graph~\cite{Goldberg2007}.
Moreover, it has been shown that a multiplicative approximation 
of antiferromagnetic Ising partition functions 
on $d$-regular graphs ($d \geq 3$) are {\sf NP}-hard~\cite{Sly}.
Compared with the complexity of these real Ising partition functions,
the imaginary Ising partition functions seem to be much more intractable.

This result also contrasts with
the recent studies on quantum computational 
complexity of Ising partition functions with 
imaginary coupling constants~\cite{Nest_circuit,Cuevas11,AharonovJones,AharonovJonesHard,AharonovTutte,Matsuo14}.
These quantum algorithms
calculate the Ising partition functions or, more generally,
Jones or Tutte polynomials
with additive error $\epsilon$ in polynomial time of $1/\epsilon$:
\begin{eqnarray*}
| \mathcal{Z} - \mathcal{Z}_{ap} | \leq \epsilon \Delta,
\end{eqnarray*}
where $\mathcal{Z}$ and $\mathcal{Z}_{ap}$ are
true and approximated values 
respectively, and $\Delta$
is a certain algorithmic scale.
Furthermore, it has been shown that
such an additive approximation is as powerful as
solving {\sf BQP}-complete problems (i.e., {\sf BQP}-hard).
This implies that
these quantum algorithms do a nontrivial task that would be intractable on a classical computer.
However, these quantum algorithms seem not to achieve an efficient multiplicative approximation,
since it is \#{\sf P}-hard as shown above.

\section{Conclusion and discussion \label{sec6}}
We have investigated {\sf IQP}
by relating it with computational complexity of Ising partition functions
with imaginary coupling constants and magnetic fields.
We found classes of {\sf IQP} 
that are classically simulatable at least in the weak sense (and almost in the strong sense).
Specifically, 
the {\sf IQP} circuits consisting only of 2D nearest-neighbor two-qubit commuting gates,
namely planar-{\sf IQP},
are classically simulatable.
However, if single-qubit rotations are allowed,
planar-{\sf IQP} becomes universal-under-postselection,
which are as powerful, with the help of postselection, as {\sf PP}.
Thus single-qubit rotations make a drastic change of
the {\sf IQP} circuits from almost strongly simulatable 
to not simulatable even in the weak sense.

The classical simulatability of
planar-{\sf IQP}
stems from the exact solvability of 
Ising models on planar lattices without magnetic fields.
Both classical computational complexity of Ising models on nonplanar lattices~\cite{Barahona,Istrail}
and quantum computation complexity of MBQC on 
nonplanar surface codes~\cite{RaussendorfNonplanar}
have been studied already.
While we did not addressed here,
computational complexity of the {\sf IQP} circuits consisting of two-qubit commuting gates 
with a nonplanar geometry is an intriguing future topic. 

By considering strong simulation of {\sf IQP},
we further explored hardness of a multiplicative approximation
of the Ising partition functions.
We have shown that a multiplicative approximation of
Ising partition functions with almost all imaginary coupling constants
is \#{\sf P}-hard even on planar lattices with a bounded-degree.

The results obtained in this work exhibit 
a rich structure of {\sf IQP},
ranging from classically simulatable
to highly intractable problems such as \#{\sf P}-hard.

\section*{Acknowledgements}
The authors thank S. Tamate for useful discussions.
KF was supported by JSPS Grant-in-Aid for Research Activity Start-up 25887034.
TM is supported by Tenure Track System by MEXT, Japan
and KAKENHI 26730003 by JSPS.

\bibliographystyle{ieeetr}
\bibliography{iqp_fujii_v3}

\end{document}